\documentclass[a4paper,american]{article}
\usepackage[T1]{fontenc}
\usepackage{array}
\usepackage{amsmath}
\usepackage{graphicx}
\usepackage{amssymb}
\usepackage{esint}
\usepackage[numbers]{natbib}
\usepackage{mciteplus}
\usepackage{psfrag}
\usepackage{enumitem}
\usepackage{babel}

\newcommand{\dsl}[1]{#1 \hspace{-5.5pt}/}

\newcommand{\cdott}{\!\cdot\!} 

\newcommand{\ie}{{\it i.e.}}

\newcommand{\kTa}{k_{1T}}
\newcommand{\kTb}{k_{2T}}
\newcommand{\kTi}{k_{iT}}
\newcommand{\AkT}{\big|\vec{k}_{T}\big|}
\newcommand{\AkTa}{\big|\vec{k}_{1T}\big|}
\newcommand{\AkTb}{\big|\vec{k}_{2T}\big|}
\newcommand{\AkTi}{\big|\vec{k}_{iT}\big|}
\newcommand{\Equation}[1]{Eq.~(\ref{#1})}

\newcommand{\ang}[2]{\langle #1#2\rangle}
\newcommand{\sqr}[2]{[ #1#2]}
\newcommand{\kTslash}{k\hspace{-5.5pt}/_T}
\newcommand{\pslash}{p\hspace{-4.5pt}/}
\newcommand{\sakT}{[3|\kTslash|a\rangle}
\newcommand{\askT}{\langle a|\kTslash|3]}

\newcommand{\inp}[2]{|#1\!\cdot\!#2|}

\usepackage{epsfig}

\newcommand{\Figure}[1]{Fig.~\ref{#1}}
\newcommand{\LipPap}{\cite{Antonov:2004hh}}
\newcommand{\LipFig}[1]{figure~#1 in~\LipPap}
\newcommand{\LipEqn}[1]{equation~(#1) in~\LipPap}

\AtBeginDocument{
  
}

\begin{document}

\title{Multi-gluon helicity amplitudes with one off-shell leg within high
energy factorization}

\author{A. van Hameren, P. Kotko, K. Kutak
\vspace{10pt} \\
\textit{\small The H.\ Niewodnicza\'nski Institute of Nuclear Physics} \\
\textit{\small Polish Academy of Sciences}\\
\textit{\small Radzikowskiego 152, 31-342 Cracow, Poland}
}

\maketitle
\begin{abstract}

Basing
 on the Slavnov-Taylor identities, we derive a new prescription to obtain gauge invariant tree-level scattering amplitudes for the process $g^{*}g\rightarrow N\, g$ within high energy factorization. Using the helicity method, we check the formalism up to several final state gluons, and we present analytical formulas for the the helicity amplitudes for $N=2$. We also compare the method with Lipatov's
effective action approach.

\end{abstract}

\section{Introduction}

The scientific plans of the Large Hadron Collider (LHC) span from
tests of complex dynamics of the Standard Model \citep{Salgado:2011wc}
to 
 searches 
for physics beyond it \citep{Albrow:2008pn}. Quantum Chromodynamics
(QCD) being a part of the Standard Model is the basic theory which
is used to 
set up the calculational background for the collisions at LHC.
 Application of perturbative
QCD relies on 
various
 factorization theorems which allow to decompose
a given process into a long-distance part 
called parton density
and a short
distance hard process.
In particular, for inclusive and jet production processes the appropriate long-distance 
part is realized in terms of various parton distribution functions.
Here we will focus on high energy factorization
\citep{Catani:1990eg} which applies when
the energy scale involved in the scattering process  is high. The evolution
equations of high energy factorization sum up logarithms of energy
accompanied by a coupling constant. Depending on the energy range and
observable one uses: the BFKL \citep{Kuraev:1977fs,Balitsky:1978ic},
BK \citep{Balitsky:1995ub,Kovchegov:1999yj}, CCFM \citep{Ciafaloni:1987ur,Catani:1989sg}
or recently proposed new evolution equations accounting for both saturation
and processes at large momentum transfers \citep{Kutak:2011fu,Kutak:2012qk}.

The important ingredients of high energy factorization are unintegrated
gluon densities, which depend not only on the longitudinal hadron momentum
fraction but on the transversal momentum as well. They have to be
convoluted with the hard process which is calculated with the initiating
gluons being off-shell. In case of processes with non-gluonic final
states, like $q$, $\bar{q}$ production for instance, the corresponding
gauge invariant amplitudes can be calculated by applying so called
high energy projectors, that resolve degrees of freedom relevant at
high energies. The more general framework given  in terms of Lipatov's action \citep{Lipatov:1995pn}
provides a prescription to calculate gauge invariant matrix elements
in a general case for any partonic process initiated by off-shell
gluons \citep{Antonov:2004hh}. For recent studies and applications, we refer to \citep{Chachamis:2012gh,Hentschinski:2011tz,Braun:2011qm,Bartels:2012mq}.

At present, various automatic numerical tools exist for the calculation of multi-gluon amplitudes for ordinary on-shell processes \citep{Mangano:2002ea,Cafarella:2007pc,Gleisberg:2008fv,Alwall:2011uj,Kleiss:2010hy}.
They operate on the amplitude level and use helicity methods, and thus are very efficient and universal.
The present study is a first step aiming at similar methods for off-shell high energy amplitudes.
We develop a new prescription for calculating gauge invariant
high energy factorizable tree-level amplitudes with one of the gluons being off-shell.
Our results turn out to be in agreement with results obtained using
Lipatov's action.

Our prescription can be summarized as follows.
It is known \citep{Leonidov:1999nc} that the ordinary Feynman graphs contributing to an amplitude in an on-shell calculation are in general not sufficient to obtain gauge invariant results if any of the external legs are off-shell.
Thus one usually
needs to consider a larger on-shell process and disentangle the required gauge invariant
off-shell amplitudes. 
 Our approach is however different: we require
the Slavnov-Taylor identities to hold for the off-shell connected
amplitudes, which in turn induces the necessary contributions which
by comparison to \citep{Leonidov:1999nc} can be interpreted as bremsstrahlung
 graphs introduced to restore gauge invariance. Our construction
holds for any number of final state gluons and as an example we give a detailed
derivation and discussion for the helicity amplitudes of the process $g^{*}g\rightarrow gg$. These
amplitudes are the dominant contribution to
forward-central di-jet production \citep{Deak:2010gk,Deak:2011ga,Deak:2009ae} and  are of particular interest for
testing parton densities at low longitudinal momentum fraction of the gluons
with a special focus on saturation effects (for overview see \citep{Kutak:2011rv}
and for recent jet studies of these effects see \citep{Kutak:2012rf}).

An extensive analyzis of tree-level multi-gluon helicity amplitudes in the high-energy limit was done also in Ref. \citep{DelDuca:1999ha}. There, the impact factors for $g^* g\rightarrow Ng$, $N\leq 3$ were disentangled at the amplitude level, however they do not correspond exactly to the kinematic situation considered in the present work.

The paper is organized as follows. Section \ref{sec:GeneralFr} is
introductory and is mainly devoted to set the notation. In particular,
we overview  high energy factorization focusing on the kinematic
setup and we introduce the framework of color
ordered helicity amplitudes which is used later in the paper. In Section
\ref{sec:Off-shell_amps} we identify the problem of gauge non-invariance
of high energy factorization amplitudes and 
 repair it using
the Slavnov-Taylor identities. In Section \ref{sec:Results} we describe
analytical and numerical tests for gauge invariant helicity amplitudes,
in particular we list the helicity amplitudes and their square for
$g^{*}g\rightarrow gg$. We also cross check our result with those
obtained by usage of Lipatov's effective action. In the appendices
we collect the color-ordered Feynman rules we needed in our construction
of gauge invariant amplitudes as well as some technicalities on the
Slavnov-Taylor identities.

\section{Prerequisites and notation}

\label{sec:GeneralFr}

\subsection{High energy factorization}

\label{sec:HighEnFact}

Our considerations of the off-shell amplitudes are embedded in the
formalism of high energy factorization. Let us thus start with a brief
recollection of this topic \citep{Catani:1990eg,Catani:1990xk,Catani:1990ka,Catani:1993ww}.

The basic observation is that the cross section for a hadronic process
can be decomposed at high energies into transversal momentum dependent
parton densities and the hard partonic cross section with off-shell
initial state partons. Thus, the intermediate object that appears
is the amplitude for the process 
\begin{equation}
g^{*}(k_{1})\, g^{*}(k_{2})\rightarrow X,\label{eq:hard_proc}\end{equation}
 where $g^{*}$ denotes an off-shell gluon, and $X$ abbreviates the
on-shell final state particles. In the considered high energy limit,
the incident momenta are para\-metrized as \begin{gather}
k_{1}^{\mu}\simeq\xi_{1}n_{a}^{\mu}+\kTa^{\mu},\label{eq:highen_mom1}\\
k_{2}^{\mu}\simeq\xi_{2}n_{b}^{\mu}+\kTb^{\mu},\end{gather}
 where $n_{a}$, $n_{b}$ are light-like momenta corresponding to
the two incident hadrons $a$ and $b$. They satisfy $n_{a}\cdott\kTi=n_{b}\cdott\kTi=0$,
and $k_{i}^{2}=\kTi^{2}=-\AkTi^{2}.$

In \citep{Catani:1990eg}, for example, heavy quark production $g^{*}(k_{1})\, g^{*}(k_{2})\to Q\,\overline{Q}$
is considered. The amplitude is calculated by taking into account
the usual tree-level Feynman graphs in the axial gauge with gauge
vector $n^{\mu}=\alpha n_{a}^{\mu}+\beta n_{b}^{\mu}$. Then the gluon
propagator is given by \begin{equation}
D^{\mu\nu}\left(k\right)=S\left(k\right)\, d^{\mu\nu}\left(k,n\right),\end{equation}
 with the scalar part\begin{equation}
S\left(k\right)=\frac{-i}{k^{2}+i\epsilon}\label{eq:scalar_prop}\end{equation}
and the transverse projector \begin{equation}
d^{\mu\nu}\left(k,n\right)=\eta^{\mu\nu}-\frac{k^{\mu}n^{\nu}+k^{\nu}n^{\mu}}{k\cdott n}+n^{2}\frac{k^{\mu}k^{\nu}}{\left(k\cdott n\right)^{2}}.\label{eq:proj_def}\end{equation}
 Above $\eta^{\mu\nu}$ is the metric tensor and we have suppressed
the color indices in $D^{\mu\nu}$. The off-shell legs include propagators,
and are contracted with eikonal couplings defined as 
\begin{gather}
\mathfrak{e}_{1}^{\mu}=\AkTa n_{a}^{\mu},\label{eq:eik1}\\
\mathfrak{e}_{2}^{\mu}=\AkTb n_{b}^{\mu}.\label{eq:eik2}\end{gather}
 It is easy to check, that 
effectively an amputated amplitude is contracted with the following vector
\begin{equation}
-i\,\mathfrak{e}_{1}^{\mu}\, D_{\mu\nu}\left(k_{1}\right)=\frac{k_{T\,1}^{\mu}}{\xi_{1}\AkTa}
\end{equation}
 and similarly for $k_{2}$, which then is shown to lead to the
correct result in the collinear limit. Furthermore, it is shown that
the result is independent of the choice of gauge-vector, proving gauge invariance.

The above might be taken as a prescription to calculate amplitudes
for arbitrary processes, but it will in general not lead to gauge-invariant
results. To achieve the latter, additional contributions are needed. 
They are 
usually obtained by considering a larger, fully on-shell, process
from which an amplitude for the off-shell process is extracted. The
situation is somewhat simpler when one of the gluons, say $k_{2}$,
reaches collinear limit, i.e.\ when $k_{2}^{2}\rightarrow0$. In
the following sections, we shall investigate such amplitudes for the
process \begin{equation}
g^{*}(k_{1})\, g(k_{2})\rightarrow g(k_{3})\ldots g(k_{N})\label{eq:process}\end{equation}
 for an arbitrary number of final state gluons.

\subsection{Color-ordered helicity amplitudes}

\label{sec:helic_color}

Let us denote our purely gluonic tree-level amplitude with $N$ external gluons
(one off-shell and $N-1$ on-shell) as $\mathcal{M}\left(\mathfrak{e}_{1},\varepsilon_{2},\ldots,\varepsilon_{N}\right)$.
For the precise definition see Section \ref{sec:Off-shell_amps}. The polarization vectors of the
on-shell gluons are denoted as $\varepsilon_{i}$. Whenever necessary,
we shall indicate the polarization state explicitly using $\varepsilon_{i}^{(\lambda)}$
with $\lambda=\pm$. Following 
e.g.\ the Kleiss-Stirling construction~\citep{Kleiss:1985yh},
the polarization vectors can be defined in terms of the gluon momentum
$k_{i}$ and an auxiliary light-like momentum $q_{i}$ satisfying
$k_{i}\cdott q_{i}\neq0$. Thus, whenever needed we use the notation
$\varepsilon_{i}^{(\lambda)}(q_{i})$ to indicate explicitly what
reference vector is used. The polarization vectors satisfy several
conditions, of which we would like to highlight two, namely \begin{gather}
\varepsilon_{i}^{(\lambda)}(q_{i})\cdott k_{i}=0,\label{eq:transversality}\\
\varepsilon_{i}^{(\lambda)}(q_{i})\cdott q_{i}=0.\label{eq:eps*q}\end{gather}
 When one changes the reference momentum, the corresponding polarization
vector transforms as
\begin{equation}
\varepsilon_{i}^{\mu}\left(q'_{i}\right)=\varepsilon_{i}^{\mu}\left(q_{i}\right)+\beta_{i}\left(q'_{i},q_{i}\right)k_{i}^{\mu},\label{eq:refmom_change}
\end{equation}
where the function $\beta_{i}$ depends on the precise representation
for the polarization vectors. It is important to note, that as long
as the amplitude $\mathcal{M}$ satisfies the Ward identity 
\begin{equation}
\mathcal{M}\left(\mathfrak{e}_{1},\varepsilon_{2},\ldots,\varepsilon_{i-1},k_{i},\varepsilon_{i+1},\ldots,\varepsilon_{N}\right)=0,\quad i=2,\dots ,N
\end{equation}
 the reference momenta can be freely changed, which has proved to
be very useful for obtaining compact expressions for multi-gluon amplitudes
(see e.g.\ \citep{Dixon:1996wi} for a review). In explicit calculations
we use the spinor representation for the polarization vectors
\begin{equation}
\varepsilon_{i}^{\mu\,\pm}\left(q_{i}\right)=\pm\frac{\left\langle q_{i}^{\mp}\right|\gamma^{\mu}\left|k_{i}^{\mp}\right\rangle }{\sqrt{2}\,\left\langle q_{i}^{\mp}|k_{i}^{\pm}\right\rangle },
\end{equation}
where
\begin{gather}
\left\langle k_{i}^{-}|k_{j}^{+}\right\rangle =\overline{u}^{-}\left(k_{i}\right)u^{+}\left(k_{j}\right)\equiv\left\langle ij\right\rangle ,\\
\left\langle k_{i}^{+}|k_{j}^{-}\right\rangle =\overline{u}^{+}\left(k_{i}\right)u^{-}\left(k_{j}\right)\equiv\left[ij\right]
\end{gather}
and the positive-energy spinors with definite helicities are defined
as $u^{\pm}\left(k_{i}\right)=\frac{1}{2}\left(1\pm\gamma_{5}\right)u\left(k_{i}\right)$. The spinor products can be explicitly calculated in the light-cone basis as
\begin{gather}
\left\langle ij\right\rangle = k_{\perp\, i}\,\sqrt{\frac{k^+_j}{k^+_i}} - k_{\perp\, j}\,\sqrt{\frac{k^+_i}{k^+_j}}, \\
\left[ ij\right] = -k^*_{\perp\, i}\sqrt{\frac{k^+_j}{k^+_i}} + k^*_{\perp\, j}\sqrt{\frac{k^+_i}{k^+_j}},
\end{gather}
where $k^{\pm} = k^0\pm k^3$ and $k_{\perp} = k^1+ik^2$.
We also use the following compact notation for spinor products with some $\dsl{\Gamma}$ matrix insertion
\begin{gather}
\left\langle q^{-}\right|\dsl{\Gamma}\left|k^{-}\right\rangle \equiv \left\langle q\right|\dsl{\Gamma}\left|k\right], \\
\left\langle q^{+}\right|\dsl{\Gamma}\left|k^{+}\right\rangle \equiv \left[ q\right|\dsl{\Gamma}\left|k\right\rangle.
\end{gather}

Another important ingredient in this respect is the use of color-ordered
amplitudes \citep{Mangano:1987xk}. They are defined via
\begin{equation}
\mathcal{M}\left(\mathfrak{e}_{1},\varepsilon_{2},\ldots,\varepsilon_{N}\right)=\sum_{\underset{{\scriptstyle \textrm{permutations}}}{\textrm{non-cyclic}}}\mathrm{Tr}\left(t^{a_{1}}\ldots t^{a_{N}}\right)\,\mathcal{A}\left(\mathfrak{e}_{1},\varepsilon_{2},\ldots,\varepsilon_{N}\right),
\label{eq:colordecom}
\end{equation}
 where the summation is over all non-cyclic permutations of the color
indices $\left\{ a_{1},\ldots,a_{N}\right\} $ corresponding to the
external gluons. The matrices $t^{a}$ are the generators of color
$\mathrm{SU}\left(N_{c}\right)$ group normalized as $\mathrm{Tr}\left( t^a t^b\right) = \delta^{ab}$. 
Note, that
the ordering of the arguments in $\mathcal{A}$ does matter and corresponds
to the order of color matrices under the trace. There are several
important properties of color-ordered amplitudes. Here we want to
mention explicitly two of them. First, gauge invariance of $\mathcal{M}$
implies gauge invariance of $\mathcal{A}$. Second, the full amplitude
squared is given by 
\begin{multline}
\sum_{\mathrm{colors}}\left|\mathcal{M}\left(\mathfrak{e}_{1},\varepsilon_{2},\ldots,\varepsilon_{N}\right)\right|^{2}  = N_{c}^{N-2}\left(N_{c}^{2}-1\right)\!\!\!\!\!\sum_{\underset{{\scriptstyle \textrm{permutations}}}{\textrm{non-cyclic}}}\left|\mathcal{A}\left(\mathfrak{e}_{1},\varepsilon_{2},\ldots,\varepsilon_{N}\right)\right|^{2}
 \\ +\mathcal{O}\left(N_{c}^{N-2}\right),\label{eq:ampl_squared}
\end{multline}
 where the contribution $\mathcal{O}\left(N_{c}^{N-2}\right)$ vanishes
for $N\leq5$.

\section{Amplitudes with one off-shell leg}

\label{sec:Off-shell_amps}

\begin{figure}
\begin{center}
\psfragscanon {\small \psfrag{1}{$k_{2}$} \psfrag{2}{$k_{1}$}
\psfrag{3}{$k_{3}$} \psfrag{4}{$k_{N}$}\includegraphics[width=3cm]{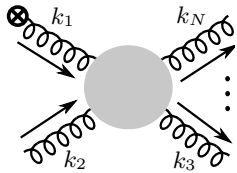}\psfragscanoff }
\end{center}
\caption{\small The color-ordered amplitude $\mathcal{A}\left(\mathfrak{e}_{1},\varepsilon_{2},\ldots,\varepsilon_{N}\right)$
defined in the main text. The special external vertex corresponds
to eikonal coupling $\mathfrak{e}_{1}^{\mu}=\AkT n_{a}^{\mu}$, while
all the other external lines are amputated and contracted with the
corresponding polarization vectors. The arrows show the choice of
the momenta signs, while the vertical dots stand for the remaining
external gluons.\label{fig:M_def}}
\end{figure}

The subject of our investigations are gluonic amplitudes with one
off-shell leg, \Equation{eq:process}. More precisely, they are obtained
from the corresponding momentum space Green function by means of the
following reduction formula\begin{multline}
\mathcal{M}\left(\mathfrak{e}_{1},\varepsilon_{2},\ldots,\varepsilon_{N}\right)=\lim_{k_{1}\cdot n_{a}\rightarrow0}\,\lim_{k_{2}^{2},\ldots,k_{N}^{2}\rightarrow0}\,\mathfrak{e}_{1}^{\mu_{1}}\varepsilon_{2}^{\mu_{2}}\ldots\varepsilon_{N}^{\mu_{N}}\,\\
S^{-1}\left(k_{2}\right)\ldots S^{-1}\left(k_{N}\right)\tilde{G}_{\mu_{1}\ldots\mu_{N}}\left(k_{1},\ldots,k_{N}\right),\label{eq:reduction}\end{multline}
where 
the
 scalar propagators are defined in Eg.~(\ref{eq:scalar_prop})
and $\tilde{G}$ is momentum space Green function defined as usual,
by Fourier transforming the vacuum matrix element of a time ordered
product of gluon fields. Note, that the first limit is a formal way
of implementing the kinematic condition (\ref{eq:highen_mom1}). The
amplitude defined in \Equation{eq:reduction} is incidentally \emph{not
gauge invariant}, see the discussion below.

Let us remark, that for ordinary, completely on-shell amplitudes the
common practice is to amputate the external gluon legs together with
potential projectors (\ref{eq:proj_def}), if one uses axial gauge
for $\tilde{G}$. In other words, in such case we can choose different
gauges for the internal (off-shell) and the external (on-shell) lines
and the scattering amplitude does not depend on this choice. It
is of course also possible to do so at intermediate stages of a calculation
when dealing with gauge-dependent quantities, however one has to be
consistent, as different gauge-dependent pieces are to eventually compose
a gauge invariant object.

In this spirit, in what follows we work in the axial gauge with $n_{a}$
being the gauge vector for internal off-shell lines (including line
$k_{1}$) while the external on-shell gluons are reduced in the Feynman
gauge, unless otherwise specified. That is, any vertex adjacent to
an external on-shell gluon is contracted with the corresponding polarization
vector directly.

In what follows, we consider a color-ordered amplitude for some specific
color ordering, say $\left(a_{1},a_{2},\ldots,a_{N}\right)$, where
$a_{i}$ is the color quantum number of the gluon with momentum $k_{i}$.
As follows from (\ref{eq:reduction}), the off-shell leg $k_{1}$
contains the full propagator and is contracted with the eikonal vertex
$\mathfrak{e}_{1}$ defined in \Equation{eq:eik1}, which we graphically
denote as a circle with a cross (\Figure{fig:M_def}). According
to Section \ref{sec:helic_color}, the amplitude under consideration
is denoted as $\mathcal{A}\left(\mathfrak{e}_{1},\varepsilon_{2},\ldots,\varepsilon_{N}\right)$.
For simplicity, we choose\begin{equation}
k_{1}^{\mu}=n_{a}^{\mu}+k_{T}^{\mu}\label{eq:kinem_k1}\end{equation}
 as the momentum of the off-shell gluon (i.e.\ we absorb $\xi$ into $n_a$ 
in \Equation{eq:highen_mom1}). 

Because of the off-shell external leg, taking into account the usual
tree-level Feynman graphs and using the usual Feynman rules will not
lead to a gauge invariant amplitude. In particular, it will not satisfy
the usual Ward identities, i.e.\ \begin{equation}
\mathcal{A}\left(\mathfrak{e}_{1},\varepsilon_{2},\ldots,\varepsilon_{i-1},k_{i},\varepsilon_{i+1},\ldots,\varepsilon_{N}\right)\neq0,\quad i=2,\ldots,N,\label{eq:ward_viol}\end{equation}
which are indispensable for the calculation of helicity amplitudes.
Indeed, it is not the Ward identity of the above form that follows
from the local gauge invariance. Rather, one should consider the non-abelian
Slavnov-Taylor (\mbox{S-T}) identites, which relate the amplitude
$\mathcal{A}$ with nonphysical polarization (i.e.\ insertion of
$k_{i}$) and the corresponding diagrams with ghosts. In the next
section, we shall wander this path to construct a new amplitude \begin{equation}
\tilde{\mathcal{A}}\left(\mathfrak{e}_{1},\varepsilon_{2},\ldots,\varepsilon_{N}\right)=\mathcal{A}\left(\mathfrak{e}_{1},\varepsilon_{2},\ldots,\varepsilon_{N}\right)+\mathcal{W}\left(\mathfrak{e}_{1},\varepsilon_{2},\ldots,\varepsilon_{N}\right),\label{eq:Aprime_def}\end{equation}
 such that \begin{equation}
\tilde{\mathcal{A}}\left(\mathfrak{e}_{1},\varepsilon_{2},\ldots,\varepsilon_{i-1},k_{i},\varepsilon_{i+1},\ldots,\varepsilon_{N}\right)=0,\quad i=2,\ldots,N.\label{eq:ward_ok}\end{equation}

Since we provide a prescription to calculate $\tilde{\mathcal{A}}\left(\mathfrak{e}_{1},\varepsilon_{2},\ldots,\varepsilon_{N}\right)$ in a specific gauge,
the fact that it satisfies Eq.~(\ref{eq:ward_ok}) is an indication that it is the correct gauge invariant result. Moreover, we show in Appendix 
\ref{sec:App_sing_behav} that (\ref{eq:Aprime_def}) posseses correct collinear and soft behaviour in the corresponding limits. 

\subsection{Restoration of gauge invariance}

\label{sec:Gauge_restor}

In what follows, we skip the first argument (corresponding to the
off-shell leg) in the color-ordered amplitude $\mathcal{A}$, i.e.\ we
write $\mathcal{A}\left(\mathfrak{e}_{1},\varepsilon_{2},\ldots,\varepsilon_{N}\right)\equiv\mathcal{A}\left(\varepsilon_{2},\ldots,\varepsilon_{N}\right)$
for compactness. In the gauge we specified above, our solution for
the `gauge restoring' amplitude $\mathcal{W}$ reads 
\begin{equation}
\mathcal{W}\left(\varepsilon_{2},\ldots,\varepsilon_{N}\right)=\left(-\frac{g}{\sqrt{2}}\right)^{N-2}\frac{-\AkT\,\varepsilon_{2}\cdott n_{a}\ldots\varepsilon_{N}\cdott n_{a}}{k_{2}\cdott n_{a}\,\left(k_{2}-k_{3}\right)\cdott n_{a}\,\dots \left(k_{2}-\ldots-k_{N-1}\right)\cdott n_{a}}.\label{eq:waste-1}
\end{equation}
The above formula can be interpreted as a straight Wilson line along $n_{a}$ with
the external gluons attached to it, however such an interpretation
is beyond the scope of this paper and shall be analyzed elsewhere.

As already mentioned, according to the S-T identities, the insertion
of $k_{i}$ generates additional gauge terms on the r.h.s of (\ref{eq:ward_viol}).
We shall see that by controlling those terms one can indeed reconstruct
$\mathcal{W}$. The S-T identities in different gauges together with
some less common issues are recollected in Appendix \ref{sub:App_ST}.

Let us consider first the (color-ordered) momentum space Green function,
from which our amplitude is obtained according to the reduction procedure
(\ref{eq:reduction}). Since we have chosen the Feynman gauge for
the external on-shell lines, the relevant S-T identity involves the
space-time derivative of the fields and in turn translates into a
momentum insertion as in (\ref{eq:ward_viol}), c.f.\ (\ref{eq:App_ST}).

\begin{figure}
\begin{raggedright} A) 
\par
\end{raggedright}
\par
\begin{centering}
\psfragscanon \psfrag{1}{$=$} \psfrag{2}{$+$} \psfrag{4}{$+\,\dots$}
\psfrag{3}{$+$} \psfrag{5}{$+\,\dots$}
\includegraphics[width=10cm]{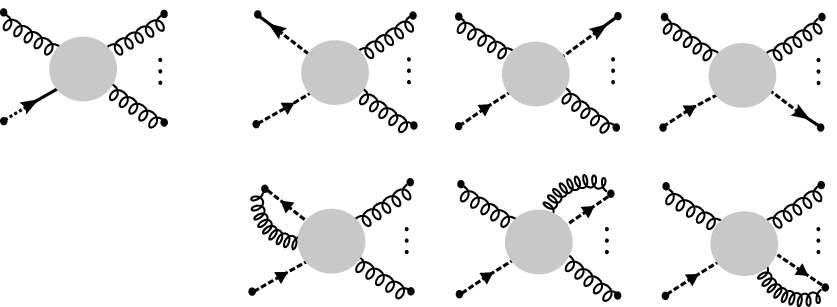}\psfragscanoff 
\par\end{centering}
\par
\begin{raggedright} B) 
\par
\end{raggedright}
\par
\begin{centering}
\psfragscanon \psfrag{1}{$=$} \psfrag{2}{$+$} \psfrag{3}{$\!\!\! =$}\includegraphics[width=10cm]{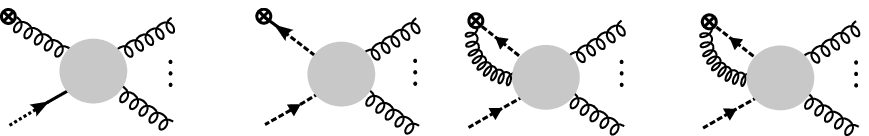}\psfragscanoff 
\par\end{centering}

\caption{{\small A) The Slavnov-Taylor identity for the Green function (with
external propagators) contracted with $k_{2}$. The contraction is
graphically denoted by the arrow transforming to a solid line (see
Appendix \ref{sub:App_color_FR}). The dashed lines with arrows are
standard notation for ghosts. Horizontal dots denote analogous diagrams
for the rest of external legs. B) Application of the identity for the amplitude for a process
with $N-1$ on-shell gluons and one off-shell leg contracted with
an eikonal vertex. Most of the gauge diagrams vanish as described
in the main text, but not all due to the off-shellness of $k_{1}$.\label{fig:ST_Green}}}
\end{figure}

The S-T identities for the Green functions are most conveniently given
in graphical form. We show an example in \Figure{fig:ST_Green}A
for the contraction with $k_{2}$. It generates a set of diagrams
involving ghosts and picks out a ghost-gluon vertex (see (\ref{eq:BRST_K})
in Appendix \ref{sub:App_ST}) without an external propagator. The
ghosts propagating to the shaded blob, which is assumed to be calculated
in the axial gauge, may need some explanation. Let us note that the ghosts
can be introduced also in the axial gauge but they decouple for ordinary
processes, see e.g.\ \citep{Leibbrandt:1987qv} and Appendix \ref{sub:App_ST}
for additional explanations. The ghost-gluon coupling is proportional
to $n_{a}^{\mu}$ and is transverse to the gluon propagator in the
axial gauge $n_{a}^{\mu}d_{\mu\nu}\left(k,n_{a}\right)=0$, thus any
internal gluon cannot be coupled to a ghost. In our case, however,
a ghost can couple to external on-shell gluons, which are taken
in the Feynman gauge and for which the projector $d_{\mu\nu}$ is
absent.

When going from the Green function to the amplitude $\mathcal{A}$
most of the additional gauge diagrams vanish (\Figure{fig:ST_Green}B),
either due to the transversality condition (\ref{eq:transversality}),
or because the residue at the physical pole is zero (due to the lack
of the external propagator, c.f. the reduction formula (\ref{eq:reduction})).
However, since $k_{1}$ is off-shell eventually one term survives,
namely the one with the ghost-gluon vertex with the amputated external
line (the second term of BRST transformation vertex (\ref{eq:BRST_K}))
contracted with the eikonal coupling. Note, that the first term in
the middle in \Figure{fig:ST_Green}B vanishes due to the relation
$k_{1}\cdott n_{a}=0$.

\begin{figure}
\begin{centering}
\psfragscanon {\small \psfrag{1}{$k_{1}$} \psfrag{2}{$k_{2}$}
\psfrag{3}{$k_{3}$} \psfrag{4}{$k_{4}$}\includegraphics[width=6.5cm]{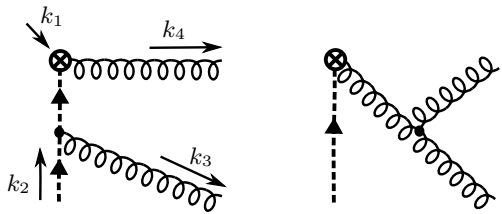}\psfragscanoff }
\par\end{centering}

\caption{\small The diagrams contributing to the Slavnov-Taylor identity for the color-ordered
amplitude $\mathcal{A}\left(\mathfrak{e}_{1},\varepsilon_{2},\varepsilon_{3},\varepsilon_{4}\right)$,
corresponding to the the process $g^{*}g\rightarrow gg$. We show
the momentum flow in the leftmost diagram only. The right diagram
is zero if the axial gauge with $n_{a}$ gauge vector is chosen. \label{fig:gauge_diag_1}}
\end{figure}

In order to illustrate the above statements, let us look at the specific
realization of the gauge diagrams from \Figure{fig:ST_Green}B
for the simplest process $g^{*}g\rightarrow gg$. We display the diagrams
in \Figure{fig:gauge_diag_1} for the color-ordered amplitude.
Since we work in the axial gauge for the internal lines, no gluon
line can attach to a ghost unless it is an external line. There is
only one possible diagram with a triple-gluon coupling in which it
is connected to the eikonal vertex (right diagram in \Figure{fig:gauge_diag_1}).
But this diagram is also zero, since we have chosen $n_{a}$ as the
gauge vector.

The remaining gauge diagrams can easily be calculated and one can
explicitly check the S-T identities (we checked it also for a general
axial gauge with gauge vector $n\neq n_{a}$). For instance, for the
left diagram from \Figure{fig:gauge_diag_1} we obtain using the
Feynman rules collected in Appendix \ref{sub:App_color_FR} \begin{equation}
\frac{g^{2}}{2}\AkT\,\frac{n_{a}\cdott\varepsilon_{3}\, n_{a}\cdott\varepsilon_{4}}{\left(k_{2}-k_{3}\right)\cdott n_{a}}\equiv\mathcal{W}_{1}\left(k_{2},\varepsilon_{3},\varepsilon_{4}\right).\label{eq:W1k2}\end{equation}
 The arguments of $\mathcal{W}_{1}$ are chosen in this way for further
convenience and show what leg is contracted with the momentum in the
corresponding S-T relation. %

\begin{figure}
\begin{centering}
\psfragscanon {\small \psfrag{1}{$\mathcal{W}_{2a}$} \psfrag{2}{$\mathcal{W}_{2b}$}
\psfrag{3}{$\mathcal{W}_{3}$}\includegraphics[width=9cm]{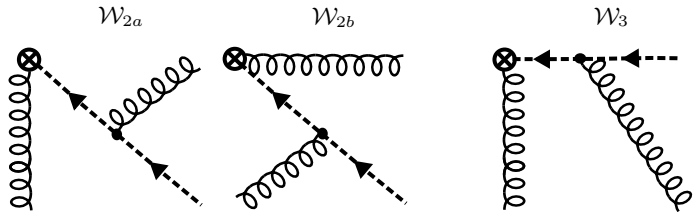}\psfragscanoff }
\par\end{centering}

\caption{\small The color-ordered gauge diagrams for the contractions of the amplitude
$\mathcal{A}\left(\varepsilon_{2},\varepsilon_{3},\varepsilon_{4}\right)$
with the external momenta; here $\mathcal{W}_{2a}$ and $\mathcal{W}_{2b}$
correspond to $\mathcal{A}\left(\varepsilon_{2},k_{3},\varepsilon_{4}\right)$,
while $\mathcal{W}_{3}$ corresponds to $\mathcal{A}\left(\varepsilon_{2},\varepsilon_{3},k_{4}\right)$.
The momentum flow is the same as in \Figure{fig:gauge_diag_1}.\label{fig:gauge_diag_2}}
\end{figure}

We can derive analogous identities involving contractions for different
external legs. The corresponding gauge diagrams are depicted in \Figure{fig:gauge_diag_2}.
Using analogous naming convention as in \Equation{eq:W1k2} they
read \begin{equation}
\mathcal{W}_{2a}\left(\varepsilon_{2},k_{3},\varepsilon_{4}\right)=\frac{g^{2}}{2}\AkT\,\frac{n_{a}\cdott\varepsilon_{2}\, n_{a}\cdott\varepsilon_{4}}{\left(k_{1}+k_{2}\right)\cdott n_{a}},\end{equation}
 \begin{equation}
\mathcal{W}_{2b}\left(\varepsilon_{2},k_{3},\varepsilon_{4}\right)=\frac{g^{2}}{2}\AkT\,\frac{n_{a}\cdott\varepsilon_{2}\, n_{a}\cdott\varepsilon_{4}}{\left(k_{1}-k_{4}\right)\cdott n_{a}},\end{equation}
 \begin{equation}
\mathcal{W}_{3}\left(\varepsilon_{2},\varepsilon_{3},k_{4}\right)=-\frac{g^{2}}{2}\AkT\,\frac{n_{a}\cdott\varepsilon_{2}\, n_{a}\cdott\varepsilon_{3}}{\left(k_{1}+k_{2}\right)\cdott n_{a}}.\end{equation}
 An immediate observation is that the gauge diagrams are proportional
to $\AkT$, thus they vanish when all the particles become on-shell;
consequently the l.h.s. of (\ref{eq:ward_viol}) equals to zero
 in that limit.

The `gauge restoring' amplitude $\mathcal{W}$ in (\ref{eq:Aprime_def})
can be obtained using the above results. First, we define $\mathcal{W}_{x}\left(\varepsilon_{2},\varepsilon_{3},\varepsilon_{4}\right)$
(note the arguments) for $x=1,2a,2b,3$ by multiplying the external
ghost lines with the longitudinal projections of the corresponding
polarization vectors. More precisely, we multiply the ghost leg $i$ with
$\varepsilon_{i}\cdott n_{a}/k_{i}\cdott n_{a}$. Hence we have \begin{equation}
\mathcal{W}_{1}\left(\varepsilon_{2},\varepsilon_{3},\varepsilon_{4}\right)=-\mathcal{W}_{3}\left(\varepsilon_{2},\varepsilon_{3},\varepsilon_{4}\right)=\frac{g^{2}}{2}\AkT\,\frac{n_{a}\cdott\varepsilon_{2}\, n_{a}\cdott\varepsilon_{3}\, n_{a}\cdott\varepsilon_{4}}{k_{2}\cdott n_{a}\, k_{4}\cdott n_{a}},\end{equation}

\begin{equation}
\mathcal{W}_{2a}\left(\varepsilon_{2},\varepsilon_{3},\varepsilon_{4}\right)=\frac{g^{2}}{2}\AkT\,\frac{n_{a}\cdott\varepsilon_{2}\, n_{a}\cdott\varepsilon_{3}\, n_{a}\cdott\varepsilon_{4}}{k_{2}\cdott n_{a}\, k_{3}\cdott n_{a}},\end{equation}
 \begin{equation}
\mathcal{W}_{2b}\left(\varepsilon_{2},\varepsilon_{3},\varepsilon_{4}\right)=-\frac{g^{2}}{2}\AkT\,\frac{n_{a}\cdott\varepsilon_{2}\, n_{a}\cdott\varepsilon_{3}\, n_{a}\cdott\varepsilon_{4}}{k_{3}\cdott n_{a}\, k_{4}\cdott n_{a}},\end{equation}
 Then, we define \begin{multline}
\mathcal{W}\left(\varepsilon_{2},\varepsilon_{3},\varepsilon_{4}\right)=\mathcal{W}_{1}\left(\varepsilon_{2},\varepsilon_{3},\varepsilon_{4}\right)+\mathcal{W}_{2a}\left(\varepsilon_{2},\varepsilon_{3},\varepsilon_{4}\right)\\
+\mathcal{W}_{2b}\left(\varepsilon_{2},\varepsilon_{3},\varepsilon_{4}\right)+\mathcal{W}_{3}\left(\varepsilon_{2},\varepsilon_{3},\varepsilon_{4}\right).\end{multline}
 Adding the diagrams, we recover the result anticipated in \Equation{eq:waste-1}
for $N=4$\begin{equation}
\mathcal{W}\left(\varepsilon_{2},\varepsilon_{3},\varepsilon_{4}\right)=-\frac{g^{2}}{2}\AkT\,\frac{n_{a}\cdott\varepsilon_{2}\, n_{a}\cdott\varepsilon_{3}\, n_{a}\cdott\varepsilon_{4}}{k_{2}\cdott n_{a}\, k_{4}\cdott n_{a}}.\end{equation}
 It is straightforward to check that indeed (\ref{eq:ward_ok}) is
satisfied with this choice of $\mathcal{W}$. It is a general
property of $\mathcal{W}$ in our choice of gauge, that the sum of
the diagrams with ghosts (and further replaced by longitudinal gluon
polarization) collapses to single terms. The complete proof of \Equation{eq:waste-1}
for any number of gluons is relegated to Appendix~\ref{sub:App_Proof_waste}.

We want to close this section with the following remark. Note that
if we choose the reference momentum for any of the polarization vectors
to be $n_{a}$, the `gauge restoring' amplitude is zero \begin{equation}
\mathcal{W}\left(\varepsilon_{2},\ldots,\varepsilon_{i-1},\varepsilon_{i}\left(n_{a}\right),\varepsilon_{i+1},\ldots,\varepsilon_{N}\right)=0\end{equation}
 for any leg $i$. It follows from the property of polarization vectors
(\ref{eq:eps*q}). Let us now note, that we can always project a polarization
vector with any reference momentum $q_{i}$ to the one with $n_{a}$
using\begin{equation}
\varepsilon_{i}^{\mu}\left(q_{i}\right)d_{\mu}^{\nu}\left(k_{i},n_{a}\right)=\varepsilon_{i}^{\nu}\left(n_{a}\right),\end{equation}
 with $d^{\mu\nu}\left(k_{i},n_{a}\right)$ defined in (\ref{eq:proj_def}).
It follows simply from the transversality condition $n_{a}\cdott d\left(k_{i},n_{a}\right)=0$.
The above remark leads to a conclusion that the gauge invariant amplitude
(\ref{eq:Aprime_def}) can be written as \begin{equation}
\tilde{\mathcal{A}}\left(\varepsilon_{2},\ldots,\varepsilon_{N}\right)=\mathcal{A}\left(\varepsilon_{2},\ldots,\varepsilon_{i-1},d\left(k_{i},n_{a}\right)\cdott\varepsilon_{i},\varepsilon_{i+1},\ldots,\varepsilon_{N}\right),\label{eq:Atil_proj}\end{equation}
 that is one can apply the transverse projector $d^{\mu\nu}\left(k_{i},n_{a}\right)$
to any number (at least one) of external on-shell legs. In particular,
one can apply the projectors to all on-shell legs. Then, (\ref{eq:ward_ok})
is trivially satisfied for any $i$, since $k_{i}\cdott d\left(k_{i},n_{a}\right)=0$
for $k_{i}^{2}=0$. Thus one could have the impression that it is also
possible for any gauge vector $n\neq n_{a}$, as the projector remains
transverse to $k_{i}$ and the `naive' Ward identity is satisfied.
This however does not mean that such amplitude is gauge invariant
-- we relegate this issue to Appendix \ref{sec:App_axial_extproj}.

\section{Analytical and numerical tests}

\label{sec:Results}

\subsection{Numerical tests}

We tested (\ref{eq:ward_ok}) numerically up to $N=12$, calculating
$\mathcal{A}$ with numerical Berends-Giele recursion~\citep{Berends:1987me}.
We also tested (\ref{eq:ward_ok}) for color-dressed amplitudes up
to $N=8$, by calculating $\mathcal{M}$ in the color-flow representation following the method of~\citep{Papadopoulos:2005ky,Duhr:2006iq}.
The color-dressed version of $\mathcal{W}$ was calculated by straightforward
multiplication with the color-dependent factor and summation over
the necessary permutations, \ie\ following \Equation{eq:colordecom} with the color-factor in the color-flow representation \citep{Kanaki:2000ms,Maltoni:2002mq}.

\subsection{Analytic expressions for $g^{*}g\rightarrow gg$}

Let us start with the amplitude with full dependence on polarization
vectors. Defining \begin{equation}
\varepsilon_{1}^{\mu}=\frac{k_{T}^{\mu}}{\AkT}\end{equation}
 we get explicitly for  color-ordering $\left(a_{1},a_{2},a_{3},a_{4}\right)$
\begin{multline}
\tilde{\mathcal{A}}\left(\varepsilon_{2},\varepsilon_{3},\varepsilon_{4}\right)=\tilde{\mathcal{A}}\left(\varepsilon_{4},\varepsilon_{3},\varepsilon_{2}\right)=\frac{g^{2}}{2\, k_{2}\cdott k_{3}k_{3}\cdott k_{4}k_{2}\cdott n_{a}k_{4}\cdott n_{a}}\\
(n_{a}\cdott\varepsilon_{3}(k_{2}\cdott k_{3}k_{3}\cdott\varepsilon_{4}k_{4}\cdott n_{a}\varepsilon_{1}\cdott\varepsilon_{2}+k_{2}\cdott n_{a}k_{3}\cdott k_{4}k_{3}\cdott\varepsilon_{2}\varepsilon_{1}\cdott\varepsilon_{4})k_{1}^{2}\\
+k_{4}\cdott n_{a}(k_{2}\cdott n_{a}((k_{2}\cdott k_{3}k_{4}\cdott\varepsilon_{3}\,\mathfrak{t}-k_{3}\cdott k_{4}(k_{2}\cdott\varepsilon_{3}\,\mathfrak{r}+2k_{2}\cdott k_{3}\varepsilon_{1}\cdott\varepsilon_{3}))\varepsilon_{2}\cdott\varepsilon_{4}\\
+k_{2}\cdott k_{3}(2(k_{1}\cdott\varepsilon_{4}k_{2}\cdott\varepsilon_{3}\varepsilon_{1}\cdott\varepsilon_{2}-k_{1}\cdott\varepsilon_{3}(k_{2}\cdott\varepsilon_{4}\varepsilon_{1}\cdott\varepsilon_{2}+k_{1}\cdott\varepsilon_{2}\varepsilon_{1}\cdott\varepsilon_{4})\\
+k_{1}\cdott\varepsilon_{2}(k_{3}\cdott\varepsilon_{4}\varepsilon_{1}\cdott\varepsilon_{3}-k_{2}\cdott\varepsilon_{3}\varepsilon_{1}\cdott\varepsilon_{4}))-(2\mathfrak{u}_{13}\cdott\varepsilon_{2}-k_{3}\cdott\varepsilon_{2}\,\mathfrak{t})\varepsilon_{3}\cdott\varepsilon_{4})\\
+k_{3}\cdott k_{4}(2k_{2}\cdott\varepsilon_{3}\,\mathfrak{s}_{24}+k_{3}\cdott\varepsilon_{2}(2\mathfrak{s}_{34}-\mathfrak{r}\,\varepsilon_{3}\cdott\varepsilon_{4})))\\
+k_{2}\cdott k_{3}((k_{4}\cdott\varepsilon_{3}n_{a}\cdott\varepsilon_{4}+k_{3}\cdott n_{a}\varepsilon_{3}\cdott\varepsilon_{4})\mathfrak{p}\cdott\varepsilon_{2}-k_{1}\cdott\varepsilon_{1}k_{1}\cdott\varepsilon_{2}k_{3}\cdott\varepsilon_{4}n_{a}\cdott\varepsilon_{3}))\\
+k_{2}\cdott n_{a}(k_{3}\cdott k_{4}(k_{2}\cdott\varepsilon_{3}n_{a}\cdott\varepsilon_{2}\varepsilon_{1}\cdott\varepsilon_{4}k_{1}^{2}-k_{1}\cdott\varepsilon_{1}k_{1}\cdott\varepsilon_{4}(k_{2}\cdott\varepsilon_{3}n_{a}\cdott\varepsilon_{2}+k_{3}\cdott\varepsilon_{2}n_{a}\cdott\varepsilon_{3}))\\
+\varepsilon_{2}\cdott\varepsilon_{3}(k_{4}\cdott n_{a}((\mathfrak{t}\, k_{1}\cdott k_{3}-2k_{3}\cdott k_{4}k_{3}\cdott\varepsilon_{1})k_{3}\cdott\varepsilon_{4}+2k_{3}\cdott k_{4}\mathfrak{u}_{13}\cdott\varepsilon_{4})\\
+k_{3}\cdott k_{4}k_{3}\cdott n_{a}\mathfrak{p}\cdott\varepsilon_{4}))+k_{1}\cdott\varepsilon_{1}\,\varepsilon_{2}\cdott n_{a}\,\varepsilon_{3}\cdott n_{a}\,\varepsilon_{4}\cdott n_{a}\, k_{2}\cdott k_{3}\, k_{3}\cdott k_{4})\end{multline}
 where we have used \begin{gather}
\mathfrak{p}^{\alpha}=k_{1}\cdott\varepsilon_{1}\, k_{1}^{\alpha}-k_{1}^{2}\varepsilon_{1}^{\alpha},\\
\mathfrak{r}=k_{1}\cdott\varepsilon_{1}+2k_{2}\cdott\varepsilon_{1}-2k_{3}\cdott\varepsilon_{1},\\
\mathfrak{s}_{ij}=k_{1}^{\alpha}\varepsilon_{1}^{\beta}\left(\varepsilon_{i\,\alpha}\varepsilon_{j\,\beta}-\varepsilon_{i\,\beta}\varepsilon_{j\,\alpha}\right),\\
\mathfrak{t}=k_{1}\cdott\varepsilon_{1}+2k_{2}\cdott\varepsilon_{1},\\
\mathfrak{u}_{ij}^{\alpha}=k_{i}^{\alpha}k_{j}\cdott\varepsilon_{1}-k_{i}\cdott k_{j}\varepsilon_{1}^{\alpha}.\end{gather}
 We have not used any particular reference momenta for the polarization
vectors -- by applying different choices the above expression can
be simplified. Note, that we are allowed to play with the reference
momenta as the above amplitude is gauge invariant; the $\mathcal{W}$
piece is already included above (the last term). The other color-ordered
amplitudes are obtained by exchanging the relevant polarization vectors
and momenta in the expression above.

Using the spinor helicity formalism one can calculate those amplitudes
explicitly for different helicity combinations. For this purpose it
proves most convenient to choose $n_{a}$ as a reference vector for
each of the on-shell gluons. Labeling spinors associated with the
light-like momenta $n_{a}$, $k_{2}$, $k_{3}$, $k_{4}$ with $a$,
$2$, $3$, $4$ respectively, and further using the notation introduced
in Section \ref{sec:helic_color} (see also~\citep{Mangano:1987xk}),
we find \begin{align}
\tilde{\mathcal{A}}(\varepsilon_{2}^{-}(n_{a}),\varepsilon_{3}^{-}(n_{a}),\varepsilon_{4}^{-}(n_{a})) & 
  = 0,\\
\tilde{\mathcal{A}}(\varepsilon_{2}^{+}(n_{a}),\varepsilon_{3}^{+}(n_{a}),\varepsilon_{4}^{+}(n_{a})) & 
  = 0,\\
\tilde{\mathcal{A}}(\varepsilon_{2}^{-}(n_{a}),\varepsilon_{3}^{-}(n_{a}),\varepsilon_{4}^{+}(n_{a})) & 
  = \frac{g^{2}}{\sqrt{2}}\,\frac{\sakT}{\AkT}\,\frac{\sqr{4}{a}^{3}}{\sqr{3}{a}\sqr{a}{2}\sqr{2}{3}\sqr{3}{4}},\\
\tilde{\mathcal{A}}(\varepsilon_{2}^{+}(n_{a}),\varepsilon_{3}^{+}(n_{a}),\varepsilon_{4}^{-}(n_{a})) & 
  = \frac{g^{2}}{\sqrt{2}}\,\frac{\askT}{\AkT}\,\frac{\ang{4}{a}^{3}}{\ang{a}{3}\ang{a}{2}\ang{2}{3}\ang{3}{4}},\\
\tilde{\mathcal{A}}(\varepsilon_{2}^{-}(n_{a}),\varepsilon_{3}^{+}(n_{a}),\varepsilon_{4}^{+}(n_{a})) & 
  = \frac{g^{2}}{\sqrt{2}}\,\frac{\askT}{\AkT}\,\frac{\ang{a}{2}^{3}}{\ang{a}{3}\ang{2}{3}\ang{3}{4}\ang{4}{a}},\\
\tilde{\mathcal{A}}(\varepsilon_{2}^{+}(n_{a}),\varepsilon_{3}^{-}(n_{a}),\varepsilon_{4}^{-}(n_{a})) & 
  = \frac{g^{2}}{\sqrt{2}}\,\frac{\sakT}{\AkT}\,\frac{\sqr{a}{2}^{3}}{\sqr{3}{a}\sqr{2}{3}\sqr{3}{4}\sqr{4}{a}},\\
\tilde{\mathcal{A}}(\varepsilon_{2}^{-}(n_{a}),\varepsilon_{3}^{+}(n_{a}),\varepsilon_{4}^{-}(n_{a})) & 
  = \frac{g^{2}}{\sqrt{2}}\,\frac{\sakT}{\AkT}\,\frac{\sqr{3}{a}^{3}}{\sqr{a}{2}\sqr{2}{3}\sqr{3}{4}\sqr{4}{a}},\\
\tilde{\mathcal{A}}(\varepsilon_{2}^{+}(n_{a}),\varepsilon_{3}^{-}(n_{a}),\varepsilon_{4}^{+}(n_{a})) & 
  = \frac{g^{2}}{\sqrt{2}}\,\frac{\askT}{\AkT}\,\frac{\ang{a}{3}^{3}}{\ang{a}{2}\ang{2}{3}\ang{3}{4}\ang{4}{a}}.\end{align}
 All momenta are assumed to be incoming in the formulas above. In
what follows we skip the reference momenta indication in the amplitudes.

We may decompose $k_{T}^{\mu}$ in terms of $k_{3}^{\mu}$ and another
light-like momentum $p^{\mu}$ following
\begin{equation}
k_{T}^{\mu}=p^{\mu}+\frac{k_{T}^{2}}{2k_{T}\!\cdot\! k_{3}}\, k_{3}^{\mu}~.
\end{equation}
Then, we have
\begin{equation}
\AkT^2=-k_{T}^{2}=2\frac{(k_{3}\!\cdot\! p)(p\!\cdot\! n_{a})}{k_{3}\!\cdot\! n_{a}}~,
\end{equation}
and
\begin{equation}
\big|\sakT\big|^{2}=\big|[3|\pslash|a\rangle\big|^{2}=4|k_{3}\!\cdot\! p||p\!\cdot\! n_{a}|~,\end{equation}
so
\begin{equation}
\left|\frac{\sakT}{\AkT}\right|^{2}=4|k_{3}\!\cdot\! p||p\!\cdot\! n_{a}|\,\frac{|k_{3}\!\cdot\! n_{a}|}{2|k_{3}\!\cdot\! p||p\!\cdot\! n_{a}|}=2|k_{3}\!\cdot\! n_{a}|~.\end{equation}
 So for the squared helicity amplitudes, we find \begin{align}
|\tilde{\mathcal{A}}(\varepsilon_{2}^{-},\varepsilon_{3}^{-},\varepsilon_{4}^{+})|^{2} & =|\tilde{\mathcal{A}}(\varepsilon_{2}^{+},\varepsilon_{3}^{+},\varepsilon_{4}^{-})|^{2}=\frac{g^{4}}{2}\,\frac{\inp{k_{4}}{n_{a}}^{4}}{\inp{n_{a}}{k_{2}}\inp{k_{2}}{k_{3}}\inp{k_{3}}{k_{4}}\inp{k_{4}}{n_{a}}}\\
|\tilde{\mathcal{A}}(\varepsilon_{2}^{-},\varepsilon_{3}^{+},\varepsilon_{4}^{+})|^{2} & =|\tilde{\mathcal{A}}(\varepsilon_{2}^{+},\varepsilon_{3}^{-},\varepsilon_{4}^{-})|^{2}=\frac{g^{4}}{2}\,\frac{\inp{k_{2}}{n_{a}}^{4}}{\inp{n_{a}}{k_{2}}\inp{k_{2}}{k_{3}}\inp{k_{3}}{k_{4}}\inp{k_{4}}{n_{a}}}\\
|\tilde{\mathcal{A}}(\varepsilon_{2}^{-},\varepsilon_{3}^{+},\varepsilon_{4}^{-})|^{2} & =|\tilde{\mathcal{A}}(\varepsilon_{2}^{+},\varepsilon_{3}^{-},\varepsilon_{4}^{+})|^{2}=\frac{g^{4}}{2}\,\frac{\inp{k_{3}}{n_{a}}^{4}}{\inp{n_{a}}{k_{2}}\inp{k_{2}}{k_{3}}\inp{k_{3}}{k_{4}}\inp{k_{4}}{n_{a}}}~.\end{align}
Applying \Equation{eq:ampl_squared} we find agreement with \citep{Deak:2009xt}. The overall factor $(k_1\cdott k_2/k_1\cdott p_2)^2$ in that publication is due to a difference in the the definition of $p_2$ there and $n_a$ here.

\subsection{Comparison with Lipatov's effective action approach}

\begin{figure}
\epsfig{figure=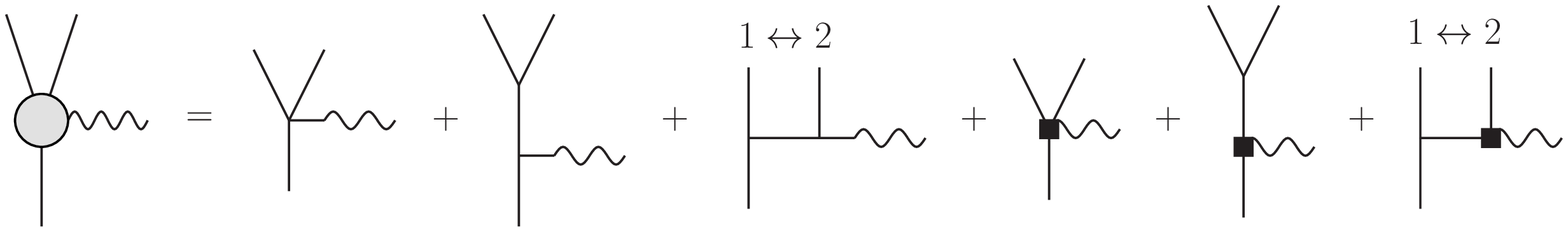,width=0.99\linewidth}
\caption{\small \label{Fig:Lipatov}The effective reggeon-gluon-gluon-gluon vertex
as obtained from \LipFig{6}, by expanding all blobs on the r.h.s\ into
graphs using \LipFig{4}.}
\end{figure}

Let us now show that the gluonic amplitudes with one off-shell leg
augmented with the `gauge restoring' amplitude (\ref{eq:waste-1})
are equivalent to the amplitudes obtained using the effective reggeon-gluon
vertices of~\LipPap. In the vertices, no external gluon is necessarily
on-shell, and the reggeon essentially is an off-shell gluon with momentum
$k_{1}^{\mu}=E(n^{+})^{\mu}+k_{T}^{\mu}$, where $(n^{+})^{\mu}=n_{a}^{\mu}/E$.
If all gluons, except the reggeon, are on-shell, the vertices correspond
to our amplitudes, modulo a factor $E/\AkT$.

This can most easily be understood considering \Figure{Fig:Lipatov}.
It graphically depicts the reggeon-gluon-gluon-gluon vertex as presented
in~\LipPap. Straight lines are gluons, and the wavy line attached
to a straight line is the off-shell gluon without a propagator, but
contracted with $\AkT^{2}(n^{+})^{\mu}=(\AkT/E)\mathfrak{e}_{1}^{\mu}$.
Here seems to be a difference with our approach, in which $\mathfrak{e}_{1}^{\mu}$
is contracted to the off-shell leg {\em including the propagator},
and in particular including the projector in the axial gauge. Notice,
however, that the collection of all graphs with a wavy line attached
to a straight line is exactly the set of usual Feynman graphs, calculated
with usual Feynman rules, and before contracting them with $\mathfrak{e}_{1}^{\mu}$
their sum forms an {\em off-shell current} $J_{1}^{\mu}$, satisfying
current conservation 
\begin{equation}
k_{1}\cdott J_{1}=0.\label{eq:curcons}\end{equation}
The off-shell current is not gauge-invariant, but \Equation{eq:curcons}
holds in any gauge. Consequently, we have 
\begin{equation}
\mathfrak{e}_{1}^{\mu}d_{\mu\nu}(k_{1},n)J_{1}^{\nu}=\mathfrak{e}_{1}\cdott J_{1}-\frac{\mathfrak{e}_{1}\cdott k_{1}}{n\cdott k_{1}}\, n\cdott J_{1}=\mathfrak{e}_{1}\cdott J_{1},\end{equation}
where the second equality follows from the fact that $\mathfrak{e}_{1}\cdott k_{1}=\AkTa n_{a}\cdott k_{1}=0$.
So we see that it does not matter for the off-shell external line
whether the projector is present or not, in any gauge.

The vertices in~\Figure{Fig:Lipatov} indicated by a square represent
the effective vertices of \LipEqn{18}, which have the essential
feature that all gluons attached to it are contracted with $(n^{+})^{\mu}$.
Since the results of~\LipPap\ are gauge-invariant, the graphs can
be calculated in the axial gauge with gauge-vector $(n^{+})^{\mu}$,
with the consequence that the last two graphs vanish, since they contain
a gluon propagator attached to the square vertex. The fourth graph
on the r.h.s.\ stays, and can readily be recognized as our `gauge
restoring' amplitude (\ref{eq:waste-1}). %
The other graphs are the usual graphs one would take into account
in a fully on-shell calculation, and are calculated following the
usual Feynman rules. The same argumentation holds for vertices with
arbitrary number of on-shell gluons. All graphs containing square
vertices vanish, except one which corresponds to (\ref{eq:waste-1}),
and the other graphs are the usual ones taken into account in a fully
on-shell calculation.

\section{Summary and outlook}

\label{sec:summary}

We provided a prescription to calculate gauge-invariant multi-gluon
scattering amplitudes within high energy factorization, for which one initial state
gluon is off-shell. Gauge-invariance was ensured by employing Slavnov-Taylor identities,
enabling the construction of the necessary extra contributions to the amplitude besides those coming from the usual Feynman graphs.
In particular, we presented compact expressions for the helicity
amplitudes of the process $g^{*}g\rightarrow gg$, and observed that they are in agreement with existing calculations. Also, we found agreement of our prescription with the effective
action approach of Lipatov.
The generalization to two off-shell legs is currently under study.

\section*{Acknowledgments}

The authors are grateful to F. Hautmann for extensive discussions
and reading the preceding notes. P.K. would like to thank also H. Arodz, L. Motyka, W.~Slominski and L. Szymanowski.
We also acknowledge useful discussions with M. Deak and H.~Jung.

During this research P.K. and K.K have been supported by NCBiR grant
LIDER/02/35/L-2/10/NCBiR/2011.

\nocite{Arodz:2010}
\nocite{Mangano:1990by}

\bibliographystyle{apsrevM}
\bibliography{offshell_amps}

\appendix

\section{Ghosts and Slavnov-Taylor identities}

\label{sub:App_ST}

Let us consider the pure Yang-Mills Lagrangian with gauge fixing and
ghost terms\begin{equation}
\mathcal{L}_{\mathrm{YM}}=-\frac{1}{4}F_{\mu\nu}^{a}F_{a}^{\mu\nu}-\frac{1}{2\lambda}\,\mathcal{F}_{a}^{2}(A)+\overline{c}_{a}\,\frac{\delta\mathcal{F}_{a}\left(A\right)}{\delta A_{\mu}^{c}}\, D_{\mu}^{cb}c_{b},\end{equation}
where $A_{a}^{\mu}$ is a gluon field, $c_{a},\,\overline{c}_{a}$
represent ghost and anti-ghost fields, further $F_{a}^{\mu\nu}=\partial^{\mu}A_{a}^{\nu}-\partial^{\nu}A_{a}^{\mu}+gf_{abc}A_{b}^{\mu}A_{c}^{\nu}$
and $D_{\mu}^{ab}=\delta^{ab}\partial_{\mu}+gf_{abc}A_{\mu}^{c}$. Here $f_{abc}$ are the usual $\mathrm{SU}\left(N_c\right)$ structure constants. 
We consider two choices of the gauge-fixing function \begin{gather}
\mathcal{F}_{a}(A)=\partial_{\mu}A_{a}^{\mu}\quad\left(\mathrm{covariant\, gauge}\right),\label{eq:App_covgauge}\\
\mathcal{F}_{a}(A)=n_{\mu}A_{a}^{\mu}\quad\left(\mathrm{general\, axial\, gauge}\right).\label{eq:App_axialgauge}\end{gather}
where in general $n^{2}\neq0$. There is still a residual freedom
in the choice of $\lambda$ parameter which can be utilized in order
to simplify calculations. 

Given the Lagrangian above, one can obtain
the Feynman rules for gluons and ghosts, see Appendix \ref{sub:App_color_FR}. 

It is usually argued that the ghosts decouple from the theory in the axial
gauge defined by $\lambda\rightarrow 0$ \citep{Leibbrandt:1987qv}. 
Indeed,
 any internal gluon line is attached to a ghost line following
\begin{equation}
n^{\mu}D_{\mu\nu}\left(k,n\right)=-i\lambda\frac{k_{\nu}}{k\cdot n}\label{eq:prop_red_lamb}\end{equation}
and thus vanishes 
 for $\lambda\rightarrow0$.
The only candidates 
 for 
non-vanishing graphs with closed ghost loops are of the type
\begin{center}
\begin{tabular}{>{\centering}m{0.88\columnwidth}>{\centering}m{0.05\columnwidth}}
$\qquad\quad$\includegraphics[width=3cm]{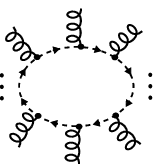}
& \addtocounter{equation}{1} 
(\theequation)
\newcounter{ghost_loop}
\setcounter{ghost_loop}{\theequation}
\end{tabular}
\end{center}
where all gluons are external.
Using the Feynman rules and dimensional regularization
for the loop integral we have (modulo couplings and phase factors)
\begin{equation}
I_{\mathrm{gh-loop}}=\int\frac{d^{D}k}{k\cdot n\left(k+p_{1}\right)\cdot n\,\ldots\left(k+p_{1}+\ldots p_{M}\right)\cdot n},
\end{equation}
where $p_{1},\ldots,p_{M}$ are the momenta of the gluons. Introducing
Feynman parameters 
and shifting the integration variable
 we get
\begin{equation}
I_{\mathrm{gh-loop}}=\int_{0}^{1}\prod_{i=1}^{M}dx_{i}\int\frac{d^{D}k\,\delta\left(1-x_{1}-\ldots-x_{M}\right)}
{\left(k\cdot n\right)^{M}}=\int\frac{d^{D}k}{\left(k\cdot n\right)^{M}}=0
\end{equation}
in dimensional regularization, so also these graphs vanish. 
In certain applications it is however convenient to consider also \emph{open}
ghost lines, and indeed the following graphs
\begin{center}
\begin{tabular}{>{\centering}m{0.88\columnwidth}>{\centering}m{0.05\columnwidth}}
$\qquad\quad$\includegraphics[width=3.5cm]{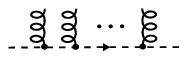}
& \addtocounter{equation}{1} 
(\theequation)
\end{tabular}
\end{center}
in which all gluons are external do not necessarily vanish. This is precisely
what is used in the present paper.

The Yang-Mills action $\mathcal{L}_{\mathrm{YM}}$ is invariant under
the Becchi-Rouet-Stora-Tyutin (BRST) transformation (see e.g.\ \citep{Arodz:2010}
for a review). The resulting Slavnov-Taylor identity for a functional
for connected Green functions $W$ reads 
\begin{multline}
\int d^{4}x\bigg\{ j_{\mu}^{a}(x)\frac{\delta}{\delta K_{\mu}^{a}(x)}-\overline{\xi}_{a}(x)\frac{\delta}{\delta H_{a}(x)}\\
+\xi_{a}(x)F^{a}\left(\frac{\delta}{\delta j(x)}\right)\bigg\} W\left[j,\xi,\overline{\xi},H,K\right]=0,\label{eq:App_ST}
\end{multline}
 where $j_{\mu}^{a}$, $\overline{\xi}_{a}$, $\xi_{a}$ are the sources
for gluons, ghosts and anti-ghosts respectively. The sources $K_{\mu}^{a}$
and $H_{a}$ generate BRST terms $\mathcal{K}_{\mu}^{a}$ and $\mathcal{H}_{a}$
respectively, where\begin{equation}
\mathcal{K}_{\mu}^{a}=\lambda\left(\partial_{\mu}c^{a}(x)-gf_{acd}A_{\mu}^{c}(x)c^{d}(x)\right),\label{eq:BRST_K}\end{equation}
 \begin{equation}
\mathcal{H}_{a}=\frac{1}{2}\lambda gf_{ade}c^{d}(x)c^{e}(x).\label{eq:BRST_H}\end{equation}
 The identities for the Green functions are derived by differentiating
\Equation{eq:App_ST} with respect to various sources. For instance,
the identity in \Figure{fig:ST_Green}A is obtained by applying
the following functional derivative \begin{equation}
\left.\frac{\delta^{n}}{\delta j_{\mu_{n}}^{a_{n}}(x_{n})\ldots\delta j_{\mu_{3}}^{a_{3}}(x_{3})\delta\xi_{a_{2}}(x_{2})\delta j_{\mu_{1}}^{a_{1}}(x_{1})}\,\left[\mathrm{Eq.}\ref{eq:App_ST}\right]\right|_{j_{\mu}^{a}=\overline{\xi}_{a}=\xi_{a}=K_{\mu}^{a}=H_{a}=0}.\end{equation}
 The various indices $x_{i},$ $a_{i}$, $\mu_{i}$ correspond to
a position, color and Lorentz index for a leg $i$. Let us note, that
depending on the form of the gauge-fixing function $F(A)$, the resulting
identities involve either the derivative of the 
 gluon field
(in covariant gauge) or contraction with the gauge vector $n$ (in
axial gauge). The former leads in momentum space to contraction with
the corresponding gluon momentum.

\section{Color-ordered Feynman rules}

\label{sub:App_color_FR}

For the reader's convenience we gather the Feynman rules for gluons and
ghosts in covariant and general axial gauges. The more customary Feynman
and pure axial gauge are obtained by setting $\lambda=1$ and $\lambda=0$
respectively.
\begin{itemize}[leftmargin=15pt]

\item for a given color ordering of external legs one writes only planar
diagrams 
\item gluon propagator 

\begin{center}
\parbox[c]{1.9cm}{%
\begin{center}
\vspace{0.3cm}
{\small \psfrag{1}[c]{$\mu$} \psfrag{2}[c]{$\nu$}\includegraphics[width=1.5cm]{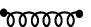} }
\end{center}}%
\parbox[c]{9cm}{%
\begin{center}
\[
=\begin{cases}
{\displaystyle \frac{-i}{k^{2}+i\varepsilon}\left[\eta^{\mu\nu}+\left(\lambda-1\right)\frac{k^{\mu}k^{\nu}}{k^{2}}\right]} & \left(\mathrm{covariant\, gauge}\right)\\
\\{\displaystyle \frac{-i}{k^{2}+i\varepsilon}\left[d^{\mu\nu}\left(k,n\right)+\lambda k^{2}\frac{k^{\mu}k^{\nu}}{\left(k\cdot n\right)^{2}}\right]} & \left(\mathrm{general\, axial\, gauge}\right)\end{cases}\]
\end{center}} 
\end{center}

\item ghost propagator (momentum flows right)

\begin{center}
\parbox[c]{1.9cm}{%
\begin{center}
\vspace{0.45cm}
\includegraphics[width=1.5cm]{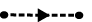} 
\end{center}}%
\parbox[c]{5cm}{%
\begin{center}
\[
=\begin{cases}
{\displaystyle \frac{-i}{k^{2}+i\epsilon}} & \left(\mathrm{covariant\, gauge}\right)\\
\\{\displaystyle \frac{-i}{k\cdott n+i\epsilon}} & \left(\mathrm{general\, axial\, gauge}\right)\end{cases}\]
\end{center}} 
\end{center}

\item triple gluon vertex (all momenta are outgoing) 

\begin{center}
\parbox[c]{2.5cm}{%
\begin{center}
\vspace{-0.3cm}
{\small \psfrag{1}[c]{$k_{1},\alpha_{1}$} \psfrag{2}{$\!\! k_{2},\alpha_{2}$}
\psfrag{3}[c]{$k_{3},\alpha_{3}$}\includegraphics[width=1.5cm]{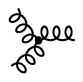} }
\end{center}}%
\parbox[c]{7.5cm}{%
\begin{center}
\begin{multline*}
=\frac{ig}{\sqrt{2}}\Big[\eta^{\alpha_{1}\alpha_{2}}\left(k_{1}-k_{2}\right)^{\alpha_{3}}+\eta^{\alpha_{2}\alpha_{3}}\left(k_{2}-k_{3}\right)^{\alpha_{1}}\\
+\eta^{\alpha_{3}\alpha_{1}}\left(k_{3}-k_{1}\right)^{\alpha_{2}}\Big]\end{multline*}
\end{center}} 
\end{center}

\item quartic gluon vertex

\begin{center}
\parbox[c]{2.5cm}{%
\begin{center}
\vspace{0.5cm}
{\small \psfrag{1}[c]{$k_{1},\alpha_{1}$} \psfrag{2}[c]{$k_{2},\alpha_{2}$}
\psfrag{3}[c,b]{$k_{3},\alpha_{3}$} \psfrag{4}[c,b]{$k_{4},\alpha_{4}$}\includegraphics[width=1.5cm]{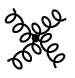} }
\end{center}}%
\parbox[c]{7.5cm}{%
\begin{center}
\[
=\frac{ig^{2}}{2}\left(2\eta^{\alpha_{1}\alpha_{3}}\eta^{\alpha_{2}\alpha_{4}}-\eta^{\alpha_{1}\alpha_{2}}\eta^{\alpha_{3}\alpha_{4}}-\eta^{\alpha_{1}\alpha_{4}}\eta^{\alpha_{2}\alpha_{3}}\right)\]
\end{center}} 
\end{center}

\item ghost-gluon vertex

\begin{center}
\parbox[c]{3cm}{%
\begin{center}
\vspace{0.5cm}
{\small \psfrag{1}[c]{$k_{1}$} \psfrag{2}{$\! k_{2},\alpha_{2}$}
\psfrag{3}[c,b]{$k_{3}$}\includegraphics[width=1.5cm]{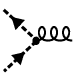} }
\end{center}}
\parbox[c]{5cm}{%
\begin{center}
\[
=\begin{cases}
\frac{-ig}{\sqrt{2}}\, k_{3}^{\alpha_{2}} & \left(\mathrm{covariant\, gauge}\right)\\
\\\frac{-ig}{\sqrt{2}}\, n^{\alpha_{2}} & \left(\mathrm{general\, axial\, gauge}\right)\end{cases}\]
\end{center}} 
\end{center}

\end{itemize}

Let us also give some auxiliary rules that are used in order to present
the Slavnov-Taylor identities (since we use color-ordered rules the
color indices are skipped):

\begin{center}
\begin{tabular}{>{\centering}m{0.4\columnwidth}>{\centering}m{0.4\columnwidth}}
\parbox[c]{1.5cm}{%
\begin{center}
{\small \psfrag{1}[c]{$\mu$} \vspace{0.5cm}
\includegraphics[width=1cm]{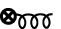} }
\end{center}%
}%
\parbox[c]{1.5cm}{%
\begin{center}
\[ =\AkT n_{a}^{\mu},\]
\end{center}}  
& 
\parbox[c]{1.2cm}{%
\begin{center}
{\small \psfrag{1}[c]{$\mu$} 
\vspace{0.5cm}
\includegraphics[width=1cm]{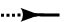} }
\end{center}}%
\parbox[c]{1.2cm}{%
\begin{center}
\[ =k^{\mu},\]
\end{center}}
\tabularnewline
\parbox[c]{1.5cm}{%
\begin{center}
{\small \psfrag{1}[c]{$\mu$} \vspace{0.5cm}
\includegraphics[width=1cm]{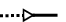} }
\end{center}}%
\parbox[c]{1.5cm}{%
\begin{center}
\[ =n^{\mu},\]
\end{center}}  
&
\parbox[c]{1.2cm}{%
\begin{center}
{\small \psfrag{1}[c]{$\mu$} \psfrag{2}[c]{$\nu$}}
\par\end{center}{\small \par}
\begin{center}
{
\vspace{0.2cm}
\includegraphics[width=1cm]{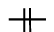} }
\end{center}}%
\parbox[c]{1.2cm}{%
\begin{center}
\[ =d^{\mu\nu}\left(k,n\right).\]
\end{center}}
\end{tabular}
\end{center}
The color-ordered Feynman rule for the BRST vertex originating in second term of Eq. (\ref{eq:BRST_K}) is 
\begin{center}
\parbox[c]{1.2cm}{%
\begin{center}
\vspace{0.5cm}
{\small \psfrag{1}[c]{$\mu$} \includegraphics[width=1.cm]{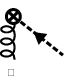} }
\end{center}}
\parbox[c]{1.5cm}{%
\begin{center}
\[
=\frac{g}{\sqrt{2}}\, n_a^{\mu}.
\]
\end{center}} 
\end{center}

\section{Axial gauge and external projectors}

\label{sec:App_axial_extproj}

It was argued in the end of Section \ref{sec:Gauge_restor} that the
gauge invariant amplitude $\tilde{\mathcal{A}}$ can be obtained by
applying at least one projector $d^{\mu\nu}\left(k_{i},n_{a}\right)$
to the external on-shell legs, cf.\ \Equation{eq:Atil_proj}. Recall
that it applies in the axial gauge with $n_{a}$ taken as the gauge vector.

One can wonder, whether the following amplitude 
\begin{equation}
\tilde{\mathcal{A}}'\left(\varepsilon_{2},\ldots,\varepsilon_{N}\right)=\mathcal{A}\left(d\left(k_{2},n\right)\cdott\varepsilon_{2},\ldots,d\left(k_{N},n\right)\cdott\varepsilon_{N}\right),\label{eq:ext_proj_1-1}
\end{equation}
 which satisfies 
\begin{equation}
\tilde{\mathcal{A}}'\left(\varepsilon_{2},\ldots,\varepsilon_{i-1},k_{i},\varepsilon_{i+1},\ldots,\varepsilon_{N}\right)=0\label{eq:Aprime_WI}
\end{equation}
 for any $i$ by definition also when $n\neq n_{a}$, is gauge invariant.
Below, we argue that it is not the case.

The point is, that although the identity (\ref{eq:Aprime_WI}) is
satisfied, the true Slavnov-Taylor identities are not fulfilled, unless
$n=n_{a}$. Actually, the S-T identities that are relevant here involve
contraction with the gauge vector $n$, not with the momentum of the
corresponding line (Appendix \ref{sub:App_ST}).

To be more specific, consider the corresponding Green function, this
time fully in the axial gauge (including the external on-shell lines)
with the gauge vector $n$. At this stage we have to keep the dependence
on the gauge parameter $\lambda$. After applying the reduction formula
(\ref{eq:reduction}) it reduces to the amplitude $\tilde{\mathcal{A}}'$.
As remarked above, the S-T identity is realized here by contracting
one of the external lines (say $k_{2}$) with $n$. Thus we have diagramatically
for the Green function (using the auxiliary Feynman rules from Appendix
\ref{sub:App_color_FR})

\begin{center}
\begin{tabular}{>{\centering}m{0.85\columnwidth}>{\centering}m{0.05\columnwidth}}
\begin{center}
\parbox[c]{2.5cm}{%
\vspace{0.4cm}
\begin{center}
\includegraphics[width=2cm]{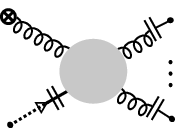}
\end{center}}
\parbox[c]{1.5cm}{
\begin{center}
\[
=-i\frac{\lambda}{k_{2}\cdot n}\times\]
\end{center}}
\parbox[c]{2.5cm}{
\vspace{0.4cm}
\begin{center}
\includegraphics[width=2cm]{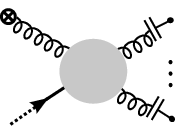}
\end{center}}
\end{center}
 & \addtocounter{equation}{1} 
(\theequation)
\end{tabular}
\end{center}

\noindent due to \Equation{eq:prop_red_lamb}. Note, that there
is no external propagator for $k_{2}$ on the r.h.s. 
(it is indicated by the lack of a dot at the end of the propagator).
On the other
hand, applying the S-T identity to the l.h.s. of this relation we get

\begin{center}
\begin{tabular}{>{\centering}m{0.85\columnwidth}>{\centering}m{0.05\columnwidth}}
\psfragscanon
\psfrag{1}[cc]{\normalsize{$\, =\lambda\times$}}
\psfrag{2}{$+\,\dots$}
\psfrag{3}[cc]{$-i\frac{\lambda}{k_2\cdot n}\times$}\includegraphics[width=6cm]{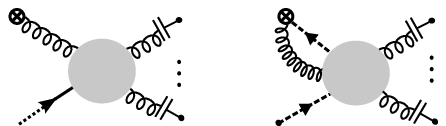}\psfragscanoff 
& \addtocounter{equation}{1} 
(\theequation)
\end{tabular}
\end{center}

\noindent where the horizontal dots denote the rest of possible gauge
terms, similar to those in \Figure{fig:ST_Green}A, but all of
them are multiplied by $\lambda$ according to (\ref{eq:BRST_K}).
Now, we apply the reduction formula to this identity. Let us note,
that there appears the factor $\lambda k_{2}^{2}/k_{2}\cdot n$ for
each term ($k_{2}^{2}$ is not cancelled on the r.h.s. as the external
propagator is $1/k_{2}\cdot n$ there), thus it can be omitted. Since
$\lambda$ cancelled, we may now set $\lambda=0$.
The final form of the S-T identity is (up to a phase factor)

\begin{center}
\begin{tabular}{>{\centering}m{0.85\columnwidth}>{\centering}m{0.05\columnwidth}}
\psfragscanon
\psfrag{1}[cc]{$=$}\includegraphics[width=5cm]{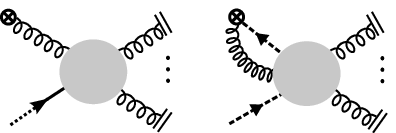}\psfragscanoff 
& \addtocounter{equation}{1} 
(\theequation)
\newcounter{ST_axial}
\setcounter{ST_axial}{\theequation}
\end{tabular}
\par\end{center}

\noindent which looks the same as \Figure{fig:ST_Green}B
with however the external projectors inserted for the on-shell gluons. 

It means that the correct identity for the amplitude with the external
projectors which is connected with the local gauge invariance is
the one above, not (\ref{eq:Aprime_WI}). Moreover, if one wants to
have zero on the r.h.s. the gauge vector has to be equal to $n_{a}$.
This is because the amplitude on the r.h.s. has the form

\begin{center}
\begin{tabular}{>{\centering}m{0.85\columnwidth}>{\centering}m{0.05\columnwidth}}
\parbox[c]{2.5cm}{
\begin{center}
\vspace{0.5cm}
\includegraphics[width=1.8cm]{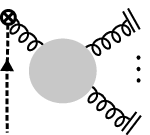}
\end{center}}
\parbox[c]{1.5cm}{%
\begin{center}
\[
\neq0\,\,\,\,\,\,\,\mathrm{if}\,\, n\neq n_{a}.\]
\end{center}} 
& \addtocounter{equation}{1} 
(\theequation)
\end{tabular}
\end{center}

\noindent The other diagrams are zero, because here both the internal
\emph{and} the external lines cannot be attached to the ghost line;
the only possibility is the diagram above. It is nonzero, because
the eikonal vertex is proportional to $n_{a}$, while $n_{a}^{\mu}d_{\mu\nu}\left(k,n\right)\neq0$.

Let us finally remark, that the relation (\arabic{ST_axial}) justifies
our assumption that one can use different gauges for on-shell and
off-shell lines in order to analyze the diagrams. We have started
with the Green function uniformly in axial gauge and obtained (\arabic{ST_axial})
which is the same as (\ref{eq:Aprime_def}) obtained in two different
gauges.

\section{Proof of the `gauge restoring' amplitude $\mathcal{W}$}

\label{sub:App_Proof_waste}

In the following we shall justify the equation (\ref{eq:waste-1}) for
any number of gluons.

The proof consist in two steps. First, let us argue that
\begin{equation}
\tilde{\mathcal{A}}\left(\varepsilon_{2},\ldots,\varepsilon_{N}\right)=\mathcal{A}\left(\varepsilon_{2},\ldots,\varepsilon_{N}\right)-\mathcal{A}\left(k_{2},\ldots,k_{N}\right)\,\alpha_{2}\ldots\alpha_{N},\label{eq:App_waste_1}
\end{equation}
 where
\begin{equation}
\alpha_{i}=\frac{\varepsilon_{i}\cdott n_{a}}{k_{i}\cdott n_{a}}.\label{eq:App_alph-1}
\end{equation}
 To this end let us show that $\tilde{\mathcal{A}}\left(\varepsilon_{2},\ldots,\varepsilon_{i-1},k_{i},\varepsilon_{i+1},\ldots,\varepsilon_{N}\right)=0$.
To be specific set $i=2$, i.e.\ replace $\varepsilon_{2}$ by $k_{2}$.
Graphically, we have (see Appendix \ref{sub:App_color_FR} for the
auxiliary Feynman rules)

\begin{center}
\begin{tabular}{>{\centering}m{0.89\columnwidth}>{\centering}m{0.05\columnwidth}}
\psfragscanon
\small
\psfrag{2}{\normalsize{$-$}}
\psfrag{1}{$\times\alpha_N$}
\psfrag{3}{\normalsize{$\! =$}}
\psfrag{5}{$\times\alpha_3$}
\psfrag{4}{$-$}
\includegraphics[width=9.5cm]{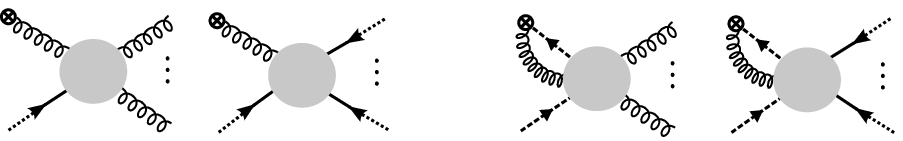}\psfragscanoff 
 & \addtocounter{equation}{1} 
(\theequation)\tabularnewline
\end{tabular}
\par\end{center}

\noindent where we have used S-T identities on the r.h.s (the rest
of the terms are zero, see the discussion to \Figure{fig:ST_Green}).
Note next, that since we use the axial gauge with $n_{a}$ as the gauge
vector for internal lines, 
only an external gluon line can couple 
to a ghost line or eikonal vertex, both of which couple via $n_{a}^{\mu}$.
Therefore we actually get zero on the r.h.s: for the first diagram
we get the factors $\varepsilon_{i}\cdott n_{a}$ for each leg, while
for the second diagram we get $k_{i}\cdott n_{i}\,\,\alpha_{i}=\varepsilon_{i}\cdott n_{a}$,
thus the result is indeed zero.

What remains is to prove that 
\begin{equation}
\mathcal{A}\left(k_{2},\ldots,k_{N}\right)\,\alpha_{2}\ldots\alpha_{n}=-\mathcal{W}\left(\varepsilon_{2},\ldots,\varepsilon_{N}\right).
\end{equation}
 It is accomplished, by noting that
\begin{equation}
\mathcal{A}\left(k_{2},\ldots,k_{N}\right)=\left(-\frac{g}{\sqrt{2}}\right)^{N-2}\AkT\frac{k_{3}\cdott n_{a}\ldots k_{N}\cdott n_{a}}{\left(k_{2}-k_{3}\right)\cdott n_{a}\dots\left(k_{2}-\ldots-k_{N-1}\right)\cdott n_{a}}
\end{equation}
 as is evident from the following diagrammatic expression

\begin{center}
\begin{tabular}{>{\centering}m{0.89\columnwidth}>{\centering}m{0.05\columnwidth}}
\psfragscanon
\small
\psfrag{2}{$k_1$}
\psfrag{3}{$k_2$}
\psfrag{4}{$k_n$}
\psfrag{5}{$k_3$}
\psfrag{1}{$\! =$}\includegraphics[width=9cm]{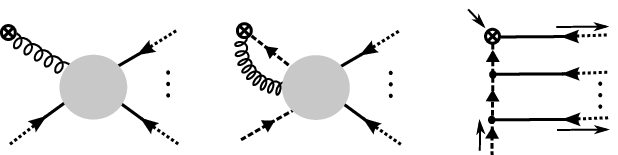}\psfragscanoff  
& \addtocounter{equation}{1} 
(\theequation)\tabularnewline
\end{tabular}
\end{center}

\noindent {where we have used the S-T identity with respect
to the leg with momentum $k_{2}$. Thus, $\mathcal{A}\left(k_{2},\ldots,k_{N}\right)$
multiplied by $\alpha_{2}\ldots\alpha_{N}$ indeed recovers (\ref{eq:waste-1}). }

\section{Singular behaviour of the amplitude $\tilde{\mathcal{A}}$}

\label{sec:App_sing_behav}

Let us show, that the gauge invariant off-shell
amplitude $\tilde{\mathcal{A}}=\mathcal{A}+\mathcal{W}$ posses the correct
singular behaviour in the collinear and soft limits. For the
purpose of this part let us supply the amplitudes with an additional
subscript denoting the number of legs, thus we write $\tilde{\mathcal{A}}\equiv\tilde{\mathcal{A}}_{N}$,
$\mathcal{A}\equiv\mathcal{A}_{N}$, $\mathcal{W}\equiv\mathcal{W}_{N}$,
if there are $N$ external legs. 

First, let us investigate the collinear limit. 
Let us start with analyzing the corresponding limit of the gauge non-invariant
amplitude $\mathcal{A}$. To this end consider the setup as in the
figure below
\begin{center}
\begin{tabular}{>{\centering}m{0.9\columnwidth}>{\centering}m{0.05\columnwidth}}
\parbox[c]{3cm}{
\begin{center}
\psfrag{1}{$k_i$}
\psfrag{2}{$k_j$}
\includegraphics[width=3cm]{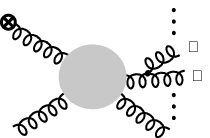}
\end{center}}
& \addtocounter{equation}{1} 
(\theequation)
\end{tabular}
\end{center}

\noindent where the momenta $k_{i}$, $k_{j}$ are parametrized as\begin{equation}
k_{i}^{\mu}=zp^{\mu}+\frac{\left|\vec{p}_{T}\right|^{2}}{2zp\cdot n_{a}}\, n_{a}^{\mu}+p_{T}^{\mu},\label{eq:App_coll_ki}\end{equation}
\begin{equation}
k_{j}^{\mu}=\left(1-z\right)p^{\mu}+\frac{\left|\vec{p}_{T}\right|^{2}}{2\left(1-z\right)p\cdot n_{a}}\, n_{a}^{\mu}-p_{T}^{\mu},\label{eq:App_coll_kj}\end{equation}
with $p^{2}=0$ and $p_{T}^{2}=-\left|\vec{p}_{T}\right|^{2}$. With
this decomposition of the momenta the scalar part of the gluon propagator
reads\begin{equation}
\frac{1}{\left(k_{i}+k_{j}\right)^{2}}=\frac{1}{\left\langle ij\right\rangle \left[ji\right]}=\frac{z\left(1-z\right)}{\left|\vec{p}_{T}\right|^{2}}.\label{eq:App_coll_prop}\end{equation}
The collinear limit is approached when \begin{equation}
\left|\vec{p}_{T}\right|\rightarrow0.\label{eq:App_coll_limit}\end{equation}
Then $k_{i}=zp$, $k_{j}=\left(1-z\right)p$ and $k_{i}+k_{j}=p$.
Let us also note the following usefull identities\begin{equation}
\varepsilon_{i}^{\left(\lambda_{i}\right)}\cdot p=-\frac{1}{z}\left(\frac{\left|\vec{p}_{T}\right|^{2}}{2zp^{+}}\,\varepsilon_{i}^{\left(\lambda_{i}\right)}\cdot n_{a}+\varepsilon_{i}^{\left(\lambda_{i}\right)}\cdot p_{T}\right),\label{eq:App_coll_eip}\end{equation}
\begin{equation}
\varepsilon_{j}^{\left(\lambda_{j}\right)}\cdot p=-\frac{1}{1-z}\left(\frac{\left|\vec{p}_{T}\right|^{2}}{2\left(1-z\right)p^{+}}\,\varepsilon_{j}^{\left(\lambda_{j}\right)}\cdot n_{a}-\varepsilon_{j}^{\left(\lambda_{j}\right)}\cdot p_{T}\right),\label{eq:App_coll_ejp}\end{equation}
which follow directly from the decompositions (\ref{eq:App_coll_ki}),
(\ref{eq:App_coll_kj}) and ortogonality relation (\ref{eq:transversality}).

In the collinear limit the amplitude behaves as\begin{equation}
\mathcal{A}_{N}\left(\varepsilon_{2},\ldots,\varepsilon_{i}^{\left(\lambda_{i}\right)},\varepsilon_{j}^{\left(\lambda_{j}\right)},\ldots,\varepsilon_{N}\right)\underset{k_{i}\parallel k_{j}}{\longrightarrow}\mathcal{A}_{N-1}\left(\varepsilon_{2},\ldots,\mathfrak{J}_{p}^{\left(\lambda_{i},\lambda_{j}\right)},\ldots,\varepsilon_{N}\right),\label{eq:App_coll_fact}\end{equation}
where the `collinear current' $\mathfrak{J}_{p}$ depends on the helicities
$\lambda_{i}$, $\lambda_{j}$ of the gluons $i,j$ and is defined as\begin{equation}
\mathfrak{J}_{p}^{\left(\lambda_{i},\lambda_{j}\right)\,\mu}=\lim_{\left|\vec{p}_{T}\right|\rightarrow0}\,\mathfrak{J}^{\left(\lambda_{i},\lambda_{j}\right)\,\mu}\left(k_{i},k_{j}\right),\end{equation}
where\begin{equation}
\mathfrak{J}^{\left(\lambda_{i},\lambda_{j}\right)\,\mu}\left(k_{i},k_{j}\right)=\frac{-id^{\mu\nu}\left(k_{i}+k_{j}\right)}{\left\langle ij\right\rangle \left[ji\right]}\, V_{3\,\nu\alpha_{i}\alpha_{j}}\left(-k_{i}-k_{j},k_{i},k_{j}\right)\varepsilon_{_{i}}^{\left(\lambda_{i}\right)\,\alpha_{i}}\varepsilon_{_{j}}^{\left(\lambda_{j}\right)\,\alpha_{j}}.\end{equation}
In the equation above we have denoted the three-point gluon vertex
as \linebreak $V_{3}^{\alpha_{1}\alpha_{2}\alpha_{3}}\left(q_{1},q_{2},q_{3}\right)$
(see Appendix \ref{sub:App_color_FR}), while the projection tensor 
is defined in Eq. (\ref{eq:proj_def}). 

We obtain\begin{multline}
\mathfrak{I}^{\left(\lambda_{i},\lambda_{j}\right)\,\mu}\left(k_{i},k_{j}\right)=\frac{\sqrt{2}g}{\left\langle ij\right\rangle \left[ji\right]}\,\\
\Bigg\{\frac{1}{1-z}\left[\varepsilon_{i}^{\left(\lambda_{i}\right)\mu}-p^{\mu}\frac{\varepsilon_{i}^{\left(\lambda_{i}\right)}\cdot n_{a}}{p\cdot n_{a}}\right]\varepsilon_{j}^{\left(\lambda_{j}\right)}\cdot p_{T}\\
+\frac{1}{z}\left[\varepsilon_{j}^{\left(\lambda_{j}\right)\mu}-p^{\mu}\frac{\varepsilon_{j}^{\left(\lambda_{j}\right)}\cdot n_{a}}{p\cdot n_{a}}\right]\varepsilon_{i}^{\left(\lambda_{i}\right)}\cdot p_{T}\\
-p_{T}^{\mu}\,\varepsilon_{i}^{\left(\lambda_{i}\right)}\cdot\varepsilon_{j}^{\left(\lambda_{j}\right)}+\mathcal{O}\left(\left|\vec{p}_{T}\right|^{2}\right)\Bigg\},\label{eq:App_coll_vertprop}\end{multline}
where we have used Eqs. (\ref{eq:App_coll_eip}), (\ref{eq:App_coll_ejp})
in order to keep track of the leading terms. In the considered limit\begin{equation}
\varepsilon_{l}^{\left(\lambda_{l}\right)}\equiv\varepsilon_{k_{l}}^{\left(\lambda_{l}\right)}=\varepsilon_{p}^{\left(\lambda_{l}\right)}+\mathcal{O}\left(\left|\vec{p}_{T}\right|\right),\,\,\,\,\, l=i,j.\label{eq:App_coll_eijlim}\end{equation}

Now, one has to fix the helicities of the gluons. For the case $\lambda_{i}=+$,
$\lambda_{j}=+$ we get the leading term\begin{equation}
\mathfrak{J}^{\left(+,+\right)\,\mu}\left(k_{i},k_{j}\right)=\sqrt{2}g\,\left(\varepsilon_{p}^{+\,\mu}-p^{\mu}\varepsilon_{p}^{+}\cdot n_{a}\right)\frac{\varepsilon_{p}^{+}\cdot p_{T}}{z\left(1-z\right)\left\langle ij\right\rangle \left[ji\right]}+\mathcal{O}\left(\left|\vec{p}_{T}\right|^{0}\right).\label{eq:App_coll_J++}\end{equation}
In order to estimate the product $\varepsilon_{p}^{+}\cdot p_{T}$
let us calculate first $\varepsilon_{p}^{+}\cdot p$ for some reference momentum $q$ and use (\ref{eq:App_coll_eip}). We get
\begin{multline}
z\varepsilon_{k_{i}}^{+}\left(q\right)\cdot p=\frac{z\left\langle q-\right|\pslash\left|k_{i}-\right\rangle }{\sqrt{2}\left\langle qk_{i}\right\rangle }=\frac{z}{\sqrt{2}}\,\frac{\left\langle qp\right\rangle \left[pk_{i}\right]}{\left\langle qk_{i}\right\rangle }\\
=\frac{z}{\sqrt{2}}\,\frac{q_{\perp}\sqrt{\frac{p\cdot n_{a}}{q\cdot n_{a}}}\,p_{\perp}^{*}z^{-\frac{1}{2}}}{q_{\perp}\sqrt{\frac{zp\cdot n_{a}}{q\cdot n_{a}}}-p_{\perp}\sqrt{\frac{q\cdot n_{a}}{zp\cdot n_{a}}}}=\frac{p_{\perp}^{*}}{\sqrt{2}}+\mathcal{O}\left(\left|\vec{p}_{T}\right|^{2}\right)\end{multline}
and thus
\begin{equation}
\varepsilon_{p}^{+}\cdot p_{T}=\frac{p_{\perp}^{*}}{\sqrt{2}}.\label{eq:App_coll_eppT}\end{equation}
However\begin{equation}
\left[ij\right]=-\frac{p_{\perp}^{*}}{\sqrt{z\left(1-z\right)}},\,\,\,\,\,\,\,\left\langle ij\right\rangle =\frac{p_{\perp}}{\sqrt{z\left(1-z\right)}}\label{eq:App_coll_[ij]}\end{equation}
so Eq. (\ref{eq:App_coll_J++}) can be written as\begin{equation}
\mathfrak{J}^{\left(+,+\right)\,\mu}\left(k_{i},k_{j}\right)=g\,\left(\varepsilon_{p}^{+\,\mu}-p^{\mu}\frac{\varepsilon_{p}^{+}\cdot n_{a}}{p\cdot n_{a}}\right)\frac{1}{\sqrt{z\left(1-z\right)}\left\langle ij\right\rangle }+\mathcal{O}\left(\left|\vec{p}_{T}\right|^{0}\right).\label{eq:App_coll_J++_1}\end{equation}
Similarily\begin{equation}
\mathfrak{J}^{\left(-,-\right)\,\mu}\left(k_{i},k_{j}\right)=g\,\left(\varepsilon_{p}^{-\,\mu}-p^{\mu}\frac{\varepsilon_{p}^{-}\cdot n_{a}}{p\cdot n_{a}}\right)\frac{1}{\sqrt{z\left(1-z\right)}\left[ji\right]}+\mathcal{O}\left(\left|\vec{p}_{T}\right|^{0}\right).\label{eq:App_coll_J--}\end{equation}
Now, consider the following helicity combination $\lambda_{i}=+$,
$\lambda_{j}=-$. We gain an additional contribution which was not
present in the previous cases, namely the one due to the last term
in (\ref{eq:App_coll_vertprop}). In order to include this, let us
decompose $p_{T}^{\mu}$ as follows\begin{equation}
p_{T}^{\mu}=\varepsilon_{p}^{+}\cdot p_{T}\left(\varepsilon_{p}^{-\,\mu}-p^{\mu}\frac{\varepsilon_{p}^{-}\cdot n_{a}}{p\cdot n_{a}}\right)+\varepsilon_{p}^{-}\cdot p_{T}\left(\varepsilon_{p}^{+\,\mu}-p^{\mu}\frac{\varepsilon_{p}^{+}\cdot n_{a}}{p\cdot n_{a}}\right).\label{eq:App_coll_pTdecomp}\end{equation}
Let us note that $\left(p_{T}^{\mu}\right)^{*}=p_{T}^{\mu}$ and $p_{T}^{2}=-2\varepsilon_{p}^{+}\cdot p_{T}\,\varepsilon_{p}^{-}\cdot p_{T}$
what gives $-\left|\vec{p}_{T}\right|^{2}$ due to (\ref{eq:App_coll_eppT})
and the fact that $\hat{p}_{T}\hat{p}_{T}^{*}=\left|\vec{p}_{T}\right|^{2}$.
Thus it is the correct representation. Therefore\begin{multline}
\mathfrak{J}^{\left(+,-\right)\,\mu}\left(k_{i},k_{j}\right)=\frac{\sqrt{2}g}{\left\langle ij\right\rangle \left[ji\right]}\,\Bigg\{\frac{z}{1-z}\left(\varepsilon_{i}^{+\mu}-p^{\mu}\frac{\varepsilon_{i}^{+}\cdot n_{a}}{p\cdot n_{a}}\right)\varepsilon_{j}^{-}\cdot p_{T}\\
+\frac{1-z}{z}\left(\varepsilon_{j}^{-\mu}-p^{\mu}\frac{\varepsilon_{j}^{-}\cdot n_{a}}{p\cdot n_{a}}\right)\varepsilon_{i}^{+}\cdot p_{T}+\mathcal{O}\left(\left|\vec{p}_{T}\right|^{2}\right)\Bigg\}\\
=g\left(\varepsilon_{p}^{+\mu}-p^{\mu}\frac{\varepsilon_{p}^{+}\cdot n_{a}}{p\cdot n_{a}}\right)\frac{z^{2}}{\sqrt{z\left(1-z\right)}\left[ji\right]}\,\\
+g\left(\varepsilon_{p}^{-\mu}-p^{\mu}\frac{\varepsilon_{p}^{-}\cdot n_{a}}{p\cdot n_{a}}\right)\frac{\left(1-z\right)^{2}}{\sqrt{z\left(1-z\right)}\left\langle ij\right\rangle }+\mathcal{O}\left(\left|\vec{p}_{T}\right|^{0}\right).\label{eq:App_coll_J+-}\end{multline}
For $\lambda_{i}=-$, $\lambda_{j}=+$ one has to replace $\varepsilon_{p}^{+}\longleftrightarrow\varepsilon_{p}^{-}$
and $\left\langle ij\right\rangle \longleftrightarrow\left[ji\right]$.
Let us note, that the coefficients of vectors in Eqs. (\ref{eq:App_coll_J++_1}), (\ref{eq:App_coll_J--}), (\ref{eq:App_coll_J+-})
are precisely the $g\rightarrow gg$ splitting amplitudes of Refs.
\citep{DelDuca:1999ha,Mangano:1990by}. Let us denote them $P_{gg}^{\left(\lambda_{p},\lambda_{i},\lambda_{j}\right)}$,
where $\lambda_{p}$ is the helicity of the gluon with momentum $p$.
Since the `collinear current' has a component proportional to $p^{\mu}$
the actual factorization does not occur for the amplitude $\mathcal{A}_{N}\left(\varepsilon_{2},\ldots,\varepsilon_{N}\right)$,
i.e. we have\begin{multline}
\mathcal{A}_{N}\left(\varepsilon_{2},\ldots\varepsilon_{i}^{\left(\lambda_{i}\right)},\varepsilon_{j}^{\left(\lambda_{j}\right)},\ldots,\varepsilon_{N}\right)\underset{k_{i}\parallel k_{j}}{\longrightarrow}\Bigg[\mathcal{A}_{N-1}\left(\varepsilon_{2},\ldots,\varepsilon_{p}^{\left(\lambda_{p}\right)},\ldots,\varepsilon_{N}\right)\\
-\mathcal{A}_{N-1}\left(\varepsilon_{2},\ldots,p,\ldots,\varepsilon_{N}\right)\,\frac{\varepsilon_{p}^{\left(\lambda_{p}\right)}\cdot n_{a}}{p\cdot n_{a}}\Bigg]\, P_{gg}^{\left(\lambda_{p},\lambda_{i},\lambda_{j}\right)}\end{multline}
with\begin{equation}
\mathcal{A}_{N-1}\left(\varepsilon_{2},\ldots,p,\ldots,\varepsilon_{N}\right)\neq0\end{equation}
due to the off-shell leg. However, as shown in the present paper the
amount of violation of gauge invariance is given by the eikonal emissions
$\mathcal{W}_{N-1}$, according to Eq. (18)\begin{equation}
\mathcal{A}_{N-1}\left(\varepsilon_{2},\ldots,p,\ldots,\varepsilon_{N}\right)=-\mathcal{W}_{N-1}\left(\varepsilon_{2},\ldots,p,\ldots,\varepsilon_{N}\right).\end{equation}
Noticing that\begin{equation}
\mathcal{W}_{N-1}\left(\varepsilon_{2},\ldots,p,\ldots,\varepsilon_{N}\right)\frac{\varepsilon_{p}\cdot n_{a}}{p\cdot n_{a}}=\mathcal{W}_{N-1}\left(\varepsilon_{2},\ldots,\varepsilon_{p},\ldots,\varepsilon_{N}\right)\label{eq:App_coll_W_N-1_rel}\end{equation}
we get\begin{equation}
\mathcal{A}_{N}\left(\varepsilon_{2},\ldots\varepsilon_{i}^{\left(\lambda_{i}\right)},\varepsilon_{j}^{\left(\lambda_{j}\right)},\ldots,\varepsilon_{N}\right)\underset{k_{i}\parallel k_{j}}{\longrightarrow}\tilde{\mathcal{A}}_{N-1}\left(\varepsilon_{2},\ldots,\varepsilon_{p}^{\left(\lambda_{p}\right)},\ldots,\varepsilon_{N}\right)P_{gg}^{\left(\lambda_{p},\lambda_{i},\lambda_{j}\right)}.\end{equation}
However \begin{equation}
\tilde{\mathcal{A}}_{N}=\mathcal{A}_{N}+\mathcal{O}\left(\left|\vec{p}_{T}\right|^{0}\right),\end{equation}
since $\mathcal{W}_{N}$ is non-singular. Therefore\begin{equation}
\tilde{\mathcal{A}}_{N}\left(\varepsilon_{2},\ldots\varepsilon_{i}^{\left(\lambda_{i}\right)},\varepsilon_{j}^{\left(\lambda_{j}\right)},\ldots,\varepsilon_{N}\right)\underset{k_{i}\parallel k_{j}}{\longrightarrow}\tilde{\mathcal{A}}_{N-1}\left(\varepsilon_{2},\ldots,\varepsilon_{p}^{\left(\lambda_{p}\right)},\ldots,\varepsilon_{N}\right)P_{gg}^{\left(\lambda_{p},\lambda_{i},\lambda_{j}\right)},\end{equation}
i.e. the amplitude $\tilde{\mathcal{A}}_{N}$ posseses the correct
collinear behaviour. 

Let us now turn to the soft limit. It is approached when the momentum
of a gluon vanishes, i.e.\begin{equation}
k_{i}^{\mu}=\lambda r^{\mu},\,\,\,\lambda\rightarrow0\end{equation}
for any fixed light-like four-vector $r^{\mu}$. In this limit the
gauge non-invariant amplitude behaves as \begin{equation}
\mathcal{A}_{N}\left(\varepsilon_{2},\ldots,\varepsilon_{i}^{\left(\lambda_{i}\right)},\varepsilon_{j}^{\left(\lambda_{j}\right)},\ldots,\varepsilon_{N}\right)\underset{\lambda\rightarrow0}{\longrightarrow}\mathcal{A}_{N-1}\left(\varepsilon_{2},\ldots,\mathfrak{I},\ldots,\varepsilon_{N}\right),\end{equation}
with the `soft current' $\mathfrak{I}^{\mu}$ defined as\begin{equation}
\mathfrak{I}^{\left(\lambda_{i},\lambda_{j}\right)\,\mu}=-\frac{1}{\lambda}\,\frac{g}{\sqrt{2}}\left(\varepsilon_{j}^{\left(\lambda_{j}\right)\,\mu}-k_{j}^{\mu}\,\frac{\varepsilon_{j}^{\left(\lambda_{j}\right)}\cdot n_{a}}{k_{j}\cdot n_{a}}\right)\,\varepsilon_{i}^{\left(\lambda_{i}\right)}\cdot\mathfrak{j}_{\mathrm{eik}}+\mathcal{O}\left(\lambda^{0}\right),\label{eq:App_coll_soft_curr}\end{equation}
where the eikonal current is\begin{equation}
\mathfrak{j}_{\mathrm{eik}}^{\mu}=\frac{k_{j}^{\mu}}{r\cdot k_{j}}.\end{equation}
We see, that the situation is similar to the collinear limit discussed
above: there is no actual factorization for the amplitude $\mathcal{A}_{N}$
since Ward identity is not satisfied due to the off-shell leg. Consequently,
the piece proportional to $k_{j}^{\mu}$ in (\ref{eq:App_coll_soft_curr})
gives a non-zero contribution. Doing, however, the same steps as for
the collinear limit (notably using Eq. (\ref{eq:App_coll_W_N-1_rel}))
we obtain
\begin{multline}
\tilde{\mathcal{A}}_{N}\left(\varepsilon_{2},\ldots,\varepsilon_{i}^{\left(\lambda_{i}\right)},\varepsilon_{j}^{\left(\lambda_{j}\right)},\ldots,\varepsilon_{N}\right)\underset{\lambda\rightarrow0}{\longrightarrow} \\
\tilde{\mathcal{A}}_{N-1}\left(\varepsilon_{2},\ldots,\varepsilon_{j}^{\left(\lambda_{j}\right)},\ldots,\varepsilon_{N}\right)\left(-\frac{1}{\lambda}\,\frac{g}{\sqrt{2}}\,\varepsilon_{i}^{\left(\lambda_{i}\right)}\cdot\mathfrak{j}_{\mathrm{eik}}\right).
\end{multline}

\end{document}